\documentclass[%
 reprint,
superscriptaddress,
 amsmath,amssymb,
 aps, 
floatfix,
]{revtex4-2}

\usepackage{booktabs}
\usepackage{multirow}

\usepackage{subcaption}

\usepackage{bm} 
\usepackage{graphicx}
\usepackage{dcolumn}
\usepackage{bm}
\usepackage{hyperref}

\begin{document}

\preprint{APS/123-QED}

\title{Strong Gravitational Lensing Effects of the Rotating Short-Haired Black Hole and Constraints from EHT Observations}

\author{Lai Zhao}

\author{Meirong Tang}

\author{Zhaoyi Xu}%
\email{zyxu@gzu.edu.cn(Corresponding author)}
\affiliation{%
 College of Physics,Guizhou University,Guiyang,550025,China
}%


\begin{abstract}

For the short hairs that have a significant impact only near the event horizon, studying their strong gravitational lensing effects is of great significance for revealing the properties of these hairs. In this study, we systematically investigated the strong gravitational lensing effects in the rotating short-haired black hole and constrained its hair parameter $Q_m$.
Specifically, \(Q_m\) causes the event horizon radius, photon - orbit radius, and impact parameter to be lower than those of the Kerr black hole. Regarding the lensing coefficients \(\bar{a}\) and \(\bar{b}\), as the spin parameter \(a\) increases, \(\bar{a}\) shows an increasing trend, while \(\bar{b}\) shows a decreasing trend.
In the observational simulations of M87* and Sgr A*, the angular position and angular separation of the relativistic image increase with the increase of \(a\), while the magnification of the image shows an opposite trend. The existence of \(Q_m\) only intensifies these trends, while parameter $k$ suppresses such tendencies.
More importantly, the rotating short-haired black hole exhibits a significant difference in time delay compared to other black hole models. Especially in the simulation of M87*, the time delay deviation between the rotating short-haired black hole and the Kerr black hole, as well as the Kerr-Newman black hole, can reach dozens of hours.
Through a comparative analysis with the observational data from the EHT, we effectively constrain the parameter space of the rotating short-haired black hole. The results indicate that this model has potential application prospects in explaining cosmic black  hole phenomena and provides a possible theoretical basis for differentiating between different black hole models.

\begin{description}
\item[Keywords]
Strong Gravitational Lensing; Rotating Short-haired Black Hole;Time Delay
\end{description}
\end{abstract}

\maketitle


\section{\label{sec:level1}Introduction}
The general theory of relativity predicts the existence of black holes(BH), but finding evidence of BH in the real universe is crucial for validating this theory. This has greatly stimulated the interest of physicists. In September 2015, the LIGO detector first captured gravitational wave signals produced by the merger of two BHs, providing strong evidence for the success of general relativity \cite{LIGOScientific:2016emj}.
In the real universe, most of the observed astrophysical BHs are rotating and fit well with the Kerr metric \cite{Teukolsky:2014vca,Tao:2023hou}. However, this does not rule out the possibility of a range of Kerr-like BHs \cite{Azreg-Ainou:2014pra,Ghosh:2014pba}.
In fact, as Xu et al. stated in the literature \cite{Xu:2021dkv}, regarding the Kerr BH as the astrophysical BH in the real universe is not entirely applicable.
This is because BH in the real universe do not exist in isolation but are coupled with dark matter or other forms of fields, making the real universe more complex. Therefore, accurately describing astrophysical BH using standard Kerr BH may present certain challenges.

If a type of Kerr-like BH solution that considers coupling with anisotropic matter could be found to describe astrophysical BH, it might be more meaningful. The short-haired  BH solution provided in the literature \cite{Brown:1997jv} better fits the real universe environment because their short-hairy BH solution considers BH solutions coupled with anisotropic matter. Their static spherically symmetric BH solution violates the no-hair theorem. The no-hair theorem states \cite{Hawking:1971vc,Israel:1967wq} that within the framework of general relativity, the properties of a BH can be completely described by three parameters: mass \( M \), charge \( Q \), and angular momentum \( J \). 
However, once Einstein's gravity is coupled with other matter, BH solutions that violate the no-hair theorem may emerge, such as \cite{Lee:2018zym,Greene:1992fw,Bizon:1990sr,Ovalle:2020kpd}.
 Therefore, studying the properties of short-hairy BH solutions \cite{Brown:1997jv} would be very interesting. However, the BH solution they provided is static and spherically symmetric. It would be more valuable to extend this solution to the rotating and axisymmetric case. This is because, on the one hand, most astrophysical BHs are rotating, so a rotating solution can better match astrophysical BHs. On the other hand, jet phenomena are common around rotating BHs, and observing these jets helps us study and understand the physical behavior near the event horizon, especially for this type of short hair that has a significant impact near the event horizon.
Based on this objective, Tang and Xu in 2022 used the Newman-Janis (NJ) algorithm to extend the short-hairy BH to the case of rotating short-hairy BH, and they also studied the impact of short-hair parameter on the BH shadow \cite{Tang:2022uwi}. Studying other properties of the rotating short-hairy BH is also very meaningful. For example, exploring their physical behavior under strong gravitational lensing can help us further understand the characteristics of rotating short-hairy BH and provide a window to test the no-hair theorem.

Gravitational lensing is an important tool for studying BHs (or other massive objects) and their surrounding environments. It has been widely applied not only in understanding the structure of spacetime (e.g., \cite{Liebes:1964zz,Mellier:1998pk,Guzik:2009cm,Schmidt:2008hc}), but also plays a crucial role in the search for dark matter (see \cite{Massey:2010hh,Vegetti:2023mgp,DiazRivero:2019hxf,Sengul:2022edu,Fairbairn:2022xln}, etc.). Meanwhile, it is an effective means to test gravitational theories (see \cite{Zhao:2024elr,Bekenstein:1993fs,Islam:2020xmy,Kumar:2020sag,Chen:2009eu}, etc.).
In a strong gravitational field, such as near a BH, when light rays approach the BH (especially near the photon sphere), the light rays will be strongly bent, and may even orbit the BH one or more times, that is, the deflection angle of the light rays exceeds \(2\pi\). In this case, shadows, photon rings, and relativistic images will appear \cite{Gralla:2019xty,Bozza:2010xqn,Bozza:2001xd,Bozza:2002af}. Regarding strong gravitational lensing, Virbhadra and Ellis analyzed the strong gravitational effects caused by a Schwarzschild BH through numerical simulations \cite{Virbhadra:1999nm}. Subsequently, Bozza \cite{Bozza:2001xd,Bozza:2002zj,Bozza:2010xqn} and Tsukamoto \cite{Tsukamoto:2016jzh}, among others, generalized it to a general static spherically symmetric spacetime. Of course, other static spherically symmetric BHs have also been studied accordingly (e.g., \cite{Man:2012ivp,Wei:2014dka,Eiroa:2013nra,Eiroa:2012fb,Eiroa:2005ag,Eiroa:2004gh,Zhao:2017cwk,Zhao:2024elr}). In addition, rotating axisymmetric BHs have also received extensive attention, such as \cite{Bozza:2008mi,Bozza:2010xqn,Bozza:2002af,Wei:2011nj,Islam:2021dyk,Islam:2021ful,Hsiao:2019ohy}.

In fact, in 2019, the EHT successfully captured images of the supermassive BH $M87^*$ \cite{EventHorizonTelescope:2019dse}, and in 2022, it captured images of the BH at the center of the Milky Way, $SgrA^*$ \cite{EventHorizonTelescope:2022wkp}. 
These breakthrough observations have triggered an upsurge of research on strong gravitational lensing.
The EHT's observations provide a new laboratory for studying the properties of BHs, such as the black hole's shadow, the jet phenomenon of the accretion disk, the event horizon, and other properties that previously could only be theoretically calculated but can now be verified through observation. Naturally, studying the strong gravitational lensing effects around BH is also of great significance, as it can magnify and distort the images of background celestial bodies, providing a unique method to detect BHs and their surrounding material distribution. Therefore, studying strong gravitational lensing in BHs that are more akin to those in the real universe holds greater physical value. The environment considered for the rotating short-haired BH \cite{Tang:2022uwi} is more similar to the real universe (anisotropic matter). Studying strong gravitational lensing in such BH can help us test the no-hair theorem, distinguish Kerr BH, and further understand the properties of short-hair near the event horizon. For this kind of hair, which has significant effects near the event horizon but is difficult to observe for distant observers, observing the behavior of light reaching near the event horizon will help us better understand the properties of this short-hair.

The paper is organized as follows: In Section \ref{sec:level2}, the short-haired BH and its event horizon information are briefly introduced. In Section \ref{sec:level3}, we calculate the lensing coefficients of the rotating short-haired BH respectively, and explore the influence of the short-hair parameter \(Q_m\) on its lensing. In Section \ref{sec:level4}, considering the rotating short-haired BH as a candidate for the supermassive BHs M87* and Sgr A*, we discuss the lensing observational effects and time delay of the rotating short-haired BH. In Section \ref{sec:level5}, we explore the use of observational data to systematically constrain the parameter space of the rotating short-haired BH. In Section \ref{sec:6}, necessary discussions are carried out. In this paper, we adopt the natural unit system, i.e., \(c = G=\hbar = 1\).

\section{\label{sec:level2}Rotating Short-Haired Black Hole}
 
In the framework of classical general relativity, the no-hair theorem states that the properties of a BH can be fully described by just three parameters: mass \( M \), charge \( Q \), and angular momentum \( J \) \cite{Hawking:1971vc,Israel:1967wq}. 
However, research indicates that the no-hair theorem might not be universal, and numerous studies have sought counterexamples.
In the 1990s, Piotr Bizon was the first to discover a BH solution with "hair" through numerical analysis, challenging the absoluteness of the no-hair theorem \cite{Bizon:1990sr}. 
Ovalle et al. constructed a spherically symmetric BH solution with hair via the gravitational decoupling method \cite{Ovalle:2020kpd}, and this was later extended to the rotating case \cite{Contreras:2021yxe}. 
Additionally, other researchers have proposed BH solutions with hair based on scalar fields or other theoretical frameworks, such as \cite{Carames:2023pde,Bakopoulos:2023hkh,Zhang:2022csi,Brihaye:2015qtu}.  In the literature \cite{Brown:1997jv}, the authors obtained short-haired BH solutions by coupling Einstein's gravity with anisotropic matter. These solutions can present de Sitter and Reissner-Nordström BHs in some cases, and in others, they yield short-haired BH solution \cite{Brown:1997jv}. The corresponding metric is given by
\begin{equation}
f(r) = 1 - \frac{2M}{r} + \frac{Q_m^{2k}}{r^{2k}}.
\label{1}
\end{equation}
Among them, \(Q_m\) is the hair strength parameter, which may represent certain quantum  effect  related hairs and has a significant impact on the structure near the event horizon.

Considering that actual BHs in the universe usually have rotational characteristics, it is particularly important to generalize such short-haired BH solutions to the rotating case. Against this background, Tang and Xu generalized the static spherically symmetric short-haired BH metric to the rotating case through the Newman-Janis (NJ) algorithm, obtained the rotating short-haired BH solution, and analyzed the influence of the short- hair parameter on the BH shadow \cite{Tang:2022uwi}. This generalization makes it possible to study the physical properties of short-haired BHs in a rotating background. Since the spacetime of a rotating short-haired BH is closer to that of the real universe, the matter distribution within it is non-vacuum and anisotropic. This model provides a basis for further exploring the properties of rotating short-haired BHs. The metric form of the rotating short-haired BH is \cite{Tang:2022uwi}
\begin{align}
ds^2& = -\left(1 - \frac{2Mr - \frac{Q_m^{2k}}{r^{2k-2}}}{\rho^2}\right) dt^2 + \frac{\rho^2}{\Delta} dr^2 \nonumber\\
&- \frac{2a \sin^2 \theta \left(2Mr - \frac{Q_m^{2 k}}{r^{2k-2}}\right)}{\rho^2} dt d\phi + \rho^2 d\theta^2   \nonumber\\
& + \frac{\Sigma \sin^2 \theta}{\rho^2} d\phi^2,
\label{2}
\end{align}
where
\begin{equation}
\rho^2 = r^2 + a^2 \cos^2 \theta ,
\label{3}
\end{equation}
\begin{equation}
\Sigma = (r^2 + a^2)^2 - a^2 \Delta ,
\label{4}
\end{equation}
\begin{equation}
 \Delta = r^2 - 2Mr + \frac{Q_m^{2 k}}{r^{2k-2}} + a^2 .
\label{5}
\end{equation}
Here, \(a\) denotes the black hole's spin parameter, and \(Q_m\) represents the hair strength parameter. As analyzed in the original literature \cite{Brown:1997jv}, the short-haired BH satisfies the Weak Energy Condition (WEC). Thus, the value of the parameter \(k\) must meet the requirement \(2k - 1\geq0\), meaning \(k\geq\frac{1}{2}\). Moreover, the trace of the energy  momentum tensor \(T = 2\rho(k - 1)>0\) (this holds when \(k > 1\)), indicating that the short-haired BH does not violate the no "short-hair" theorem \cite{Nunez:1996xv,Acharya:2024kvv}. When \(k = 1\), the metric (\ref{2}) simplifies to the classical Kerr-Newman BH. When \(k>1\), the metric (\ref{2}) depicts a short-haired BH. In the following discussion, we will focus on analyzing the short-haired BH models corresponding to parameters $k=1$, $3/2$, and $k=2$.

\begin{figure*}[]
\includegraphics[width=1 \textwidth]{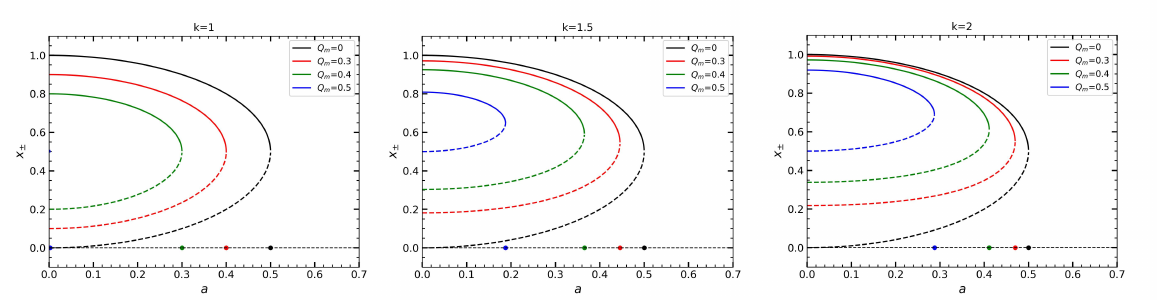}
\caption{
The event horizons and Cauchy horizons under different parameter conditions. The solid lines represent the event horizons \( x_{+} \) , and the dashed lines represent the Cauchy horizons \( x_{-} \).
}
\label{fig1}
\end{figure*}

To facilitate subsequent analysis, we've performed a non-dimensionalization of the metric (\ref{2}), using \(2M\) as the unit to non-dimensionalize physical quantities (e.g., \(r\rightarrow2Mx\)). In this framework, for the parameter models under discussion, the event horizon of the rotating short-haired BH can be determined by the condition \(g^{rr}=0\), which is equivalent to \(\Delta = 0\).
As shown in Figure \ref{fig1}, we illustrate the distribution characteristics of the event horizon for the rotating short-haired BH, where the black curve represents the event horizon of the  Kerr BH. Evidently, the introduction of the hair parameter $Q_m$ significantly alters the positional distribution of the event horizon. For different parameters $k$, the analysis shows that as the value of $k$ increases, the event horizon radius gradually converges toward the Kerr BH event horizon. This indicates that in the large $k$ value limit, the characteristics of the rotating short-haired BH progressively approach those of the Kerr BH. 
Compared to the  Kerr BH, the event horizon radius of the rotating short-haired BH is always smaller. This phenomenon suggests that the introduction of the hair parameter not only changes the spacetime structure of the BH but also provides the possibility to study the physical properties of the hair parameter in strong gravitational field environments, thereby offering a new theoretical perspective for understanding the characteristics of the rotating short-haired BH.

\section{\label{sec:level3}Strong Gravitational Lensing in the Spacetime of the Rotating Short-haired Black Hole}

For the metric of a rotating BH, to facilitate the calculations, we confine the light rays to the equatorial plane, i.e., \( \theta = \pi/2 \). The spacetime line element can be written as
\begin{equation}
ds^2 = -A(x) dt^2 + B(x) dx^2 + C(x) d\phi^2 - D(x) dt d\phi .
\label{10}
\end{equation}
By corresponding the metric of the rotating short-haired BH (\ref{2}) with equation (\ref{10}), we can obtain
\begin{align}
 A(x) &= \left(1 - \frac{x - \frac{Q_m^{2k}}{x^{2k-2}}}{x^2} \right) ,\\
 B(x) &= \frac{x^2}{\Delta}, \\
 C(x) &= \frac{\Sigma}{x^2} ,\\
 D(x) &= \frac{2a \left( x - \frac{Q_m^{2k}}{x^{2k-2}} \right)}{x^2}.
\label{11}
\end{align}

For the light rays confined to the equatorial plane, the corresponding Lagrangian is
\begin{equation}
L = \frac{1}{2} g_{\mu\nu} \frac{dx^\mu}{d\lambda} \frac{dx^\nu}{d\lambda},
\label{15}
\end{equation}
where \(\lambda\) is the affine parameter. Due to the axisymmetry of the rotating short-hairy BH, there are two conserved quantities: energy and angular momentum, given by
\begin{equation}
E = \frac{\partial L}{\partial \dot{t}} = -g_{tt} \dot{t} - g_{t\phi} \dot{\phi},
\label{16}
\end{equation}
\begin{equation}
L = -\frac{\partial L}{\partial \dot{\phi}} = g_{t\phi} \dot{t} + g_{\phi\phi} \dot{\phi},
\label{17}
\end{equation}
where the overdot represents differentiation with respect to the affine parameter, i.e., \( d/d\lambda \). When choosing an appropriate affine parameter such that \( E = 1 \), the equations of motion for light can be written as
\begin{equation}
\dot{t} = \frac{4C - 2DL}{4AC + D^2}, 
\label{18}
\end{equation}
\begin{equation}
 \dot{\phi} = \frac{2D + 4AL}{4AC + D^2}, 
\label{19}
\end{equation}
\begin{equation}
\dot{x} = \pm 2 \sqrt{\frac{C - DL - AL^2}{B(4AC + D^2)}}. 
\label{20}
\end{equation}
The effective potential can be defined through equation (\ref{20}) as
\begin{equation}
 V_{\text{eff}}(x) = -\dot{x}^2 = -4 \frac{C - DL - AL^2}{B(4AC + D^2)}. 
\label{21}
\end{equation}
By using the impact parameter \( u = L/E = L \) to replace the angular momentum \( L \), equation (\ref{21}) can be written as
\begin{equation}
V_{\text{eff}}(x) = -4 \frac{C - Du - Au^2}{B(4AC + D^2)}. 
\label{22}
\end{equation}
In the above expressions, \( A \), \( B \), \( C \), and \( D \) are all functions of \( x \).

According to the effective potential, when light rays from infinity approach a rotating short-haired BH, the light rays will be deflected near a specific radius \(x_0\). Some of the deflected light rays escape to infinity and can be detected by distant observers. When light rays approach the specific orbits of the BH, that is, the photon orbits, these orbits are highly unstable. A slight perturbation can cause the light rays to either be gravitationally captured and fall into the black hole or escape to infinity. The existence of photon orbits determines the boundary of the BH shadow and provides the conditions for the formation of the lensing effect.

\begin{figure*}[]
\includegraphics[width=1 \textwidth]{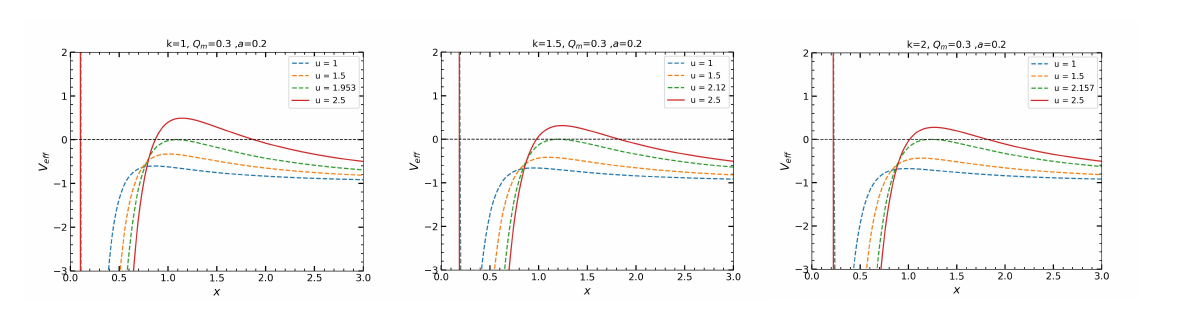}
\caption{
The effective potential under different BH models, where the intersection point of the green dashed line and the black dashed line represents the position of unstable photon orbits. Here we take $Q_m=0.3$, $a=0.2$, and from left to right correspond to BH models with $k=1$, $1.5$, $2$.}
\label{b}
\end{figure*}

\begin{figure*}[]
\includegraphics[width=1 \textwidth]{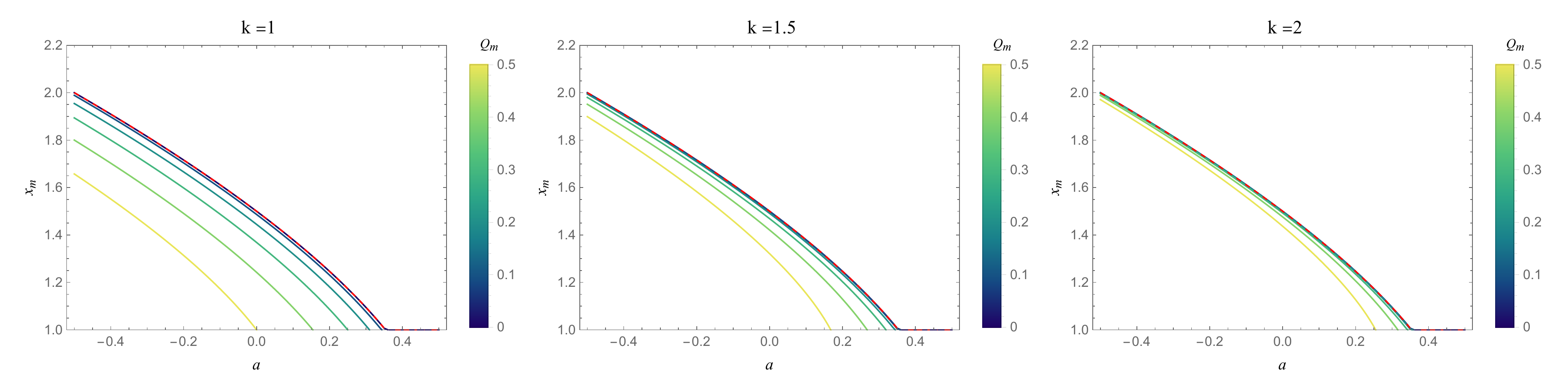}
\caption{
The variation of unstable photon orbit radius with the spin parameter a and hair parameter under different BH models, from left to right corresponding to BH models with $k=1$, $1.5$, and $2$. The red dashed line in the figure represents the Kerr BH case.}
\label{c}
\end{figure*}

The condition for an unstable Photon orbit is given by \cite{Harko:2009xf}

\begin{equation}
V_{\text{eff}}(x)=0,\; \left. \frac{dV_{\text{eff}}(x)}{dx} \right|_{x_m} = 0, \; \left. \frac{d^2 V_{\text{eff}}(x)}{dx^2} \right|_{x_m} < 0.
\label{23}
\end{equation}
Combining equations (\ref{22}) and (\ref{23}), the orbit equation for the unstable photon orbit can be derived as
\begin{equation}
A(x)C'(x) - A'(x)C(x) + u \left( A'(x)D(x) - A(x)D'(x) \right) = 0.
\label{24}
\end{equation}
Generally, the above equation has multiple solutions. Only the largest one is outside the event horizon radius, so it is defined as the radius of the unstable photon orbit. This can be clearly seen from the graph of the effective potential. As shown in Figure \ref{b}, when the impact parameter reaches the critical value (the green dashed line), the radius of the corresponding unstable photon orbit is marked as \(x = x_m\). In the following discussion, we'll use \(x_m\) to represent the radius of the unstable photon orbit.
As presented in Figure \ref{c}, because of the hair parameter \(Q_m\), the radius \(x_m\) of the unstable photon orbit differs significantly among various models. As the spin parameter increases, \(x_m\) in different models gradually drops. 
Furthermore, under the same parameter conditions, the $x_m$ value of a rotating short-haired BH is smaller than the corresponding value for a Kerr BH (\(Q_m = 0\)), and as the parameter $k$ increases, the $x_m$ value of the rotating short-haired BH gradually approaches the corresponding value of the Kerr BH. This indicates that the parameter $k$ has the effect of suppressing the hair parameter effect, and when the $k$ value is large, the behavior of the rotating short-haired BH will degenerate to that of the Kerr BH. The red dashed line in the figure represents the Kerr BH case.

When a light ray incident from infinity with a certain impact parameter \(u\) reaches the vicinity of the BH at \(x_0\). At this position, the radial velocity of the light ray is zero, while the angular velocity is non-zero. At this time, the light ray is symmetrically deflected to infinity. Since the radial velocity of the light ray is zero at the shortest distance \(x_0\), the corresponding effective potential reaches its extreme value and satisfies the condition \(V_{eff}(x)=0\) (as shown in Figure \ref{b}). Through this condition, the relationship between the impact parameter \(u\) and the shortest distance \(x_0\) can be deduced from equation(\ref{22}) as
\begin{align}
 L& = u(x_0) = \frac{-D(x_0) + \sqrt{4A(x_0)C(x_0) + D(x_0)^2}}{2A(x_0)}\nonumber \\
 &= \frac{aQ_m^{2k} x^{2-2k} - ax + x^2 \sqrt{a^2 + Q_m^{2k} x^{2-2k} + x^2 - x}}{Q_m^{2k} x^{2-2k} + x^2 - x}. 
\label{25}
\end{align}

\begin{figure*}[]
\includegraphics[width=1 \textwidth]{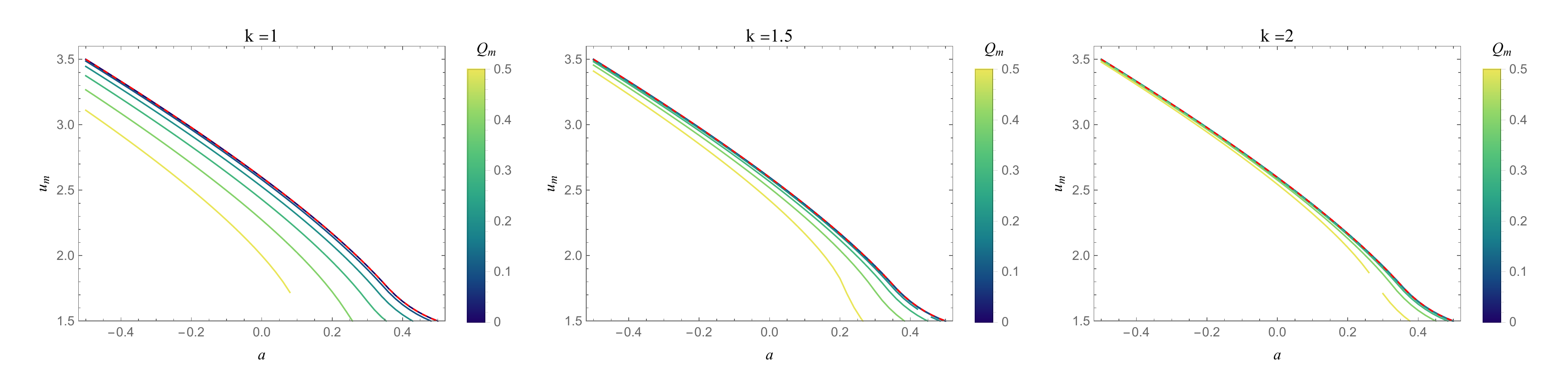}
\caption{
The variation of impact parameter with the spin parameter $a$ and hair parameter under different BH models, from left to right corresponding to BH models with $k=1$, $1.5$, and $2$. The red dashed line in the figure represents the Kerr BH case.}
\label{d}
\end{figure*}

In the above equation, \(x_0\) represents the shortest distance that the light ray reaches the BH. Since we are mainly concerned with the behavior of light rays near the unstable photon orbit, that is, the case of \(x_0\approx x_m\), the impact parameter corresponding to the unstable photon orbit can be expressed as \(u(x_m)\). The relationship between this impact parameter and the spin parameter \(a\) and the hair parameter \(Q_m\) can be visually presented in Figure \ref{d}. As the spin parameter \(a\) or the short hair parameter \(Q_m\) increases, the impact parameter \(u(x_m)\) gradually decreases, and this trend is consistent with the change trend of the photon-orbit radius.
Here, we define the photon's orbiting direction as counter clockwise. For the case of \(a > 0\) (the 	BH rotates counter  clockwise), the photon's orbiting direction is the same as the spin direction of the BH, and such an orbit is called a prograde orbit. Conversely, when \(a < 0\), the photon's orbiting direction is opposite to the spin direction of the black hole, which is called a retrograde orbit. The radius of the photon's prograde orbit is usually smaller than that of the retrograde orbit. This is because the spin effect enhances the stability of the photon closer to the BH on the prograde orbit through gravitational dragging. While on the retrograde orbit, the reverse dragging effect of the spin makes it more difficult for the photon to approach the BH, so the orbit radius is larger (see Figure \ref{c}).

In the strong field limit, the deflection angle \( \alpha_D \) can be given by \cite{Bozza:2002zj,Bozza:2002af}
\begin{equation}
\alpha_D(x_0) = I(x_0) - \pi,
\label{26}
\end{equation}
where
\begin{align}
I(x_0) &= 2 \int_{x_0}^\infty \frac{d\phi}{dx} dx \nonumber \\
&= 2 \int_{x_0}^\infty  \frac{\sqrt{A_0 B} (2Au + D)}{\sqrt{4AC + D^2} \sqrt{A_0 C - AC_0 + u(AD_0 - A_0 D)}} dx.
\label{27}
\end{align}
For the convenience of writing, it should be noted here that all capital letters in this paper represent functions of \(x\), and all letters with the subscript “0” correspond to functions when \(x = x_0\).

To handle the integral (\ref{27}), we draw on the approaches presented in the references \cite{Bozza:2002zj,Tsukamoto:2016jzh}. Specifically, we expand the deflection angle in the vicinity of the unstable photon orbit radius. Additionally, we utilize an intermediate quantity re-defined by scholar Naoki Tsukamoto under the strong  field limit to reconstruct the integral range. The form of this variable can be found in the definition given in \cite{Tsukamoto:2016jzh}, and its expression is
\begin{equation}
z = 1 - \frac{x_0}{x}.
\label{28}
\end{equation}
This variable has been well applied in corresponding literature, such as \cite{Fu:2021fxn, Islam:2021dyk}. By using the above intermediate variable, equation (\ref{27}) becomes
\begin{equation}
I(x_0) = \int_0^1 R(z, x_0) f(z, x_0) dz ,
\label{29}
\end{equation}
where
\begin{equation}
R(z, x_0) = \frac{2x^2}{x_0} \frac{\sqrt{B} (2A_0 Au + A_0 D)}{\sqrt{CA_0} \sqrt{4AC + D^2}} ,
\label{30}
\end{equation}
\begin{equation}
 f(z, x_0) = \frac{1}{\sqrt{A_0 - \frac{AC_0}{C} + \frac{u}{C} (AD_0 - A_0 D)}} .
\label{31}
\end{equation}
Evaluating the above expressions, we find that \( R(z, x_0) \) is positive definite everywhere from 0 to 1. However, \( f(z, x_0) \) diverges as \( \displaystyle\lim_{z \rightarrow 0}f(z, x_0) \). To avoid this issue, we can expand the denominator in a series up to the second-order term. Here, we define the expansion function as \( K(z, x_0) = A_0 - \frac{A C_0}{C} + \frac{u}{C} (A D_0 - A_0 D) \), then equation (\ref{31}) can be approximated as
\begin{equation}
f_0(z, x_0) = \frac{1}{\sqrt{\gamma_0(x_0) + \gamma_1(x_0) z + \gamma_2(x_0) z^2}} ,
\label{32}
\end{equation}
where
\begin{widetext}
\begin{equation}
\gamma_0(x_0) = K(z, x_0) |_{z=0} = 0,
\label{33}
\end{equation}

\begin{align}
\gamma_1(x_0) = \left. \frac{\partial K(z, x_0)}{\partial z} \right|_{z=0} = x_0 \left( u D_0 - C_0 \right) \left( \frac{A_0' C_0 - A_0 C_0'}{C_0^2} \right) - u A_0 x_0 \left( \frac{D_0' C_0 - D_0 C_0'}{C_0^2} \right) ,
\label{34}
\end{align}

\begin{align}
\gamma_2(x_0) &= \frac{1}{2!} \left. \frac{\partial^2 K(z, x_0)}{\partial z^2} \right|_{z=0} = \frac{1}{2} \left[ x_0^2 \left( u D_0 - C_0 \right) \left( \frac{A_0'' C_0 - A_0 C_0'' C_0^2 - 2 C_0 C_0' (A_0' C_0 - A_0 C_0')}{C_0^4} \right) \right. \nonumber\\
&\left.
- u A_0 x_0^2 \left( \frac{D_0'' C_0 - D_0 C_0'' C_0^2 - 2 C_0 C_0' (D_0' C_0 - D_0 C_0')}{C_0^4} \right) \right] .
\label{35}
\end{align}
\end{widetext}
In the above equations, the symbol $’$ denotes the first derivative, and $''$ denotes the second derivative.

According to the description of the strong - field limit, when light approaches the vicinity of the photon  orbit radius \(x_m\), the deflection angle of the light increases sharply.
At this point, the deflection angle can be described by an analytical expansion as \cite{Bozza:2002zj,Bozza:2002af,Tsukamoto:2016jzh}
\begin{equation}
\alpha_D(b) = -\bar{a} \log\left(\frac{\theta D_{OL}}{u_m} - 1\right) + \bar{b} + O((b - b_m) \log(b - b_m)) .
\label{36}
\end{equation}
The angular separation between the image and the lens can be approximated as \( \theta \approx \frac{u}{D_{OL}} \), where \( D_{OL} \) is the distance from the observer to the lens. The corresponding lens coefficients in equation (\ref{36}) are
\begin{equation}
\bar{a} = \frac{R(0, x_m)}{2 \sqrt{\gamma_2(x_m)}} ,
\label{37}
\end{equation}
and
\begin{equation}
\bar{b} = -\pi + I_R(x_m) + \bar{a} \log\left(\frac{c x_0^2}{u_m^2}\right),
\label{38}
\end{equation}
where 
\begin{equation}
I_R(x_0) = \int_0^1 \left( R(z, x_0) f(z, x_0) - R(0, x_m) f_0(z, x_0) \right) dz ,
\label{39}
\end{equation}
and 
and \( c \) is the expansion coefficient of the impact parameter at \( u_m \):
\begin{equation}
u - u_m = c (x_0 - x_m)^2.
\label{40}
\end{equation}

\begin{figure*}[]
\includegraphics[width=1 \textwidth]{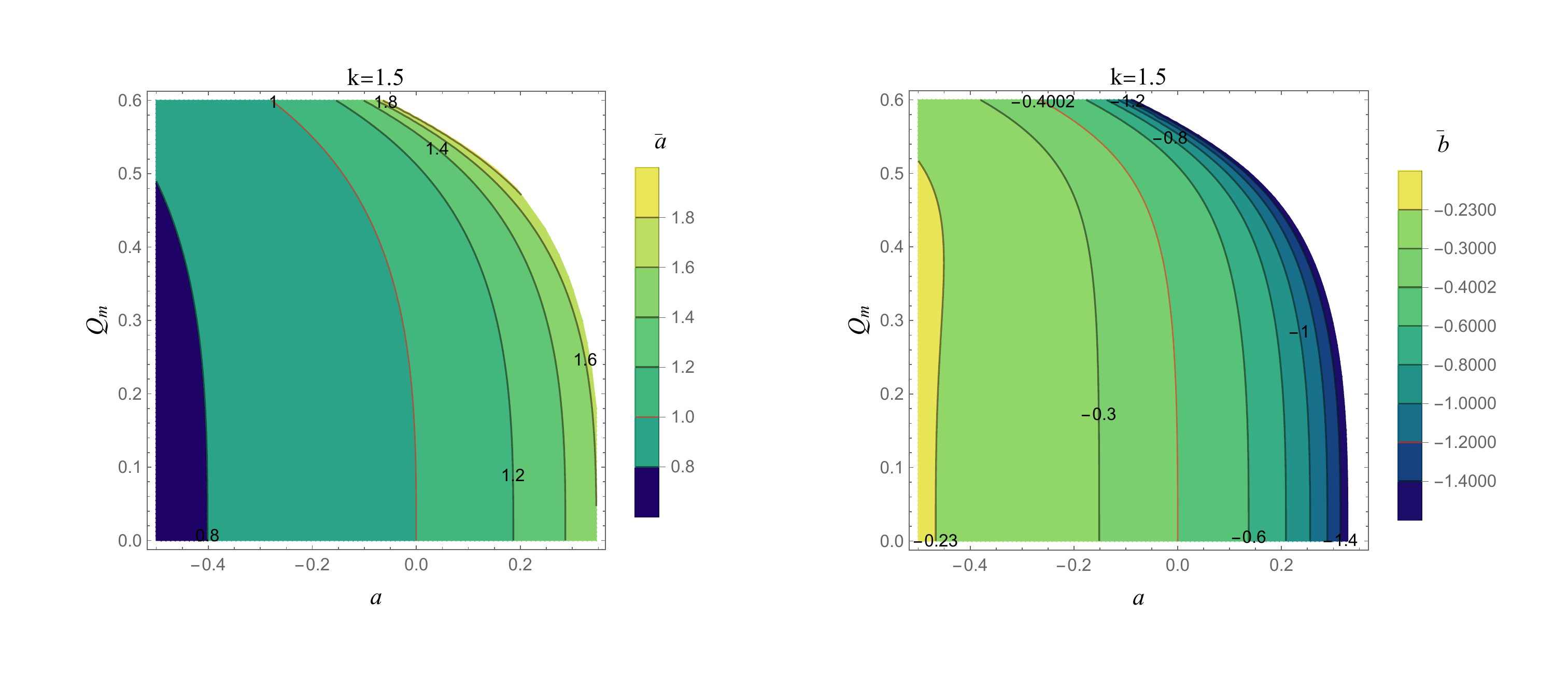}
\caption{
The variation of the lensing coefficients $\bar{a}$ and $\bar{b}$ with the spin parameter $a$ and the hair parameter $Q_m$. Here, $k = 1.5$ corresponds to the case of the rotating short-haired BH.}
\label{e}
\end{figure*}

Theoretically, substituting the corresponding expressions should yield an analytical expression for the deflection angle. However, due to the excessive complexity and length of the expression, we have instead plotted the relationships among the lensing coefficients, the short hair parameter, and the spin parameter. As shown in Figure \ref{e}, the deflection coefficient $\bar{a}$ increases with the increase of the spin parameter $a$, and the presence of the hair parameter makes this growth more rapid. When the hair parameter vanishes, i.e., $Q_m = 0$, the rotating short-haired BH degenerates into a Kerr BH. The behavior of the deflection coefficient $\bar{b}$ is exactly opposite to that of the coefficient $\bar{a}$, and the presence of  hair causes the corresponding coefficient to decrease more rapidly.
When the parameters $a = 0$ and $Q_m = 0$, the short-haired BH degenerates into a Schwarzschild BH. At this point, the lensing coefficient $\bar{a}=1$ and $\bar{b}=-0.4002$, which is in good agreement with the lensing coefficient values of the Schwarzschild BH \cite{Bozza:2002zj}(see the intersections of the red contour lines and the abscissa in the left and right plots of Figure \ref{e}, as well as Table \ref{table1} ). When the parameters $a = 0$, $k = 1$, and $Q_m\neq0$, the short-haired BH degenerates into a Reissner-Nordström (RN) BH. Our calculated results are in excellent agreement with those calculated by Bozza \cite{Bozza:2002zj} (Table \ref{table1}).
Furthermore, from the data in Table \ref{table1}, it can be clearly observed that under the same conditions, as the parameter $k$ value increases, its corresponding coefficient values gradually approach the parameter values of the Kerr BH ($Q_m=0$). This indicates that the increase in parameter $k$ weakens the influence of the hair parameter, causing the properties of the rotating short-haired BH to gradually approach those of the Kerr BH.

\begin{figure*}[]
\includegraphics[width=1 \textwidth]{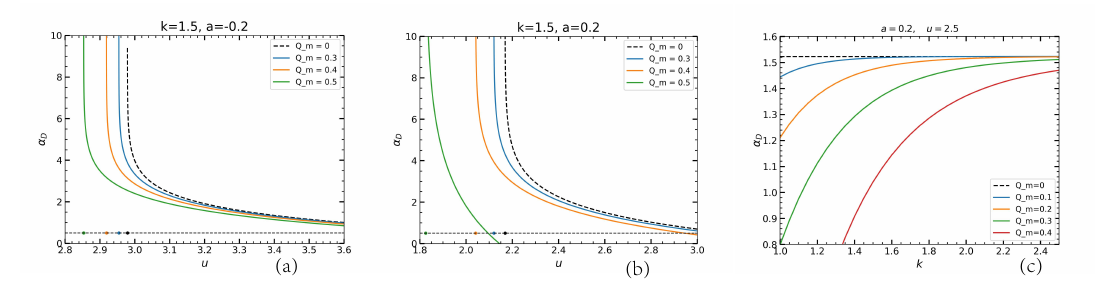}
\caption{
Deflection angle characteristics of rotating short-haired BHs in the strong deflection limit. (a) The variation of deflection angle with impact parameter $u$ under different hair parameters, with spin parameter $a=-0.2$ and hair parameter $k=1.5$; (b) The variation of deflection angle with impact parameter $u$ under different hair parameters, with spin parameter $a=0.2$ and hair parameter $k=1.5$; (c) The relationship between deflection angle and hair parameter $k$ for different hair parameter values, with spin parameter $a=0.2$ and impact parameter $u=2.5$. The black dashed line represents the Kerr BH case.}
\label{f}
\end{figure*}

In addition, in Figure \ref{f}, we have plotted the variation of the deflection angle of the short-haired BH with the impact parameter under different values of the hair parameter and the spin parameter. Evidently, as the impact parameter $u$ continuously decreases, different hair parameters always correspond to different divergence points (the points on the dashed lines in Figure \ref{f} correspond to the values of the impact parameter at divergence). Among them, the deflection angle of the Kerr BH is greater than that of the rotating short-haired BH under the same impact parameter (the black dashed line in Figure \ref{f} represents the case of the Kerr BH).
In general, the introduction of the hair parameter leads to divergent characteristics in the deflection angle at smaller impact parameters. Notably, both the lensing coefficients $\bar{a}$ and $\bar{b}$, as well as the deflection angle $\alpha$, exhibit changes that show similarity to the standard Kerr BH. Further observation reveals that as the parameter $k$ increases, the trends of these physical quantities gradually converge with the behavior of the Kerr BH. Additional analysis indicates that when $k$ takes larger values, the physical characteristics of rotating short-haired BHs will tend to degenerate with those of Kerr BH (Figure \ref{f}c), thereby reducing the feasibility of distinguishing between these two types of BHs through gravitational lensing effects. Therefore, from an observational perspective, smaller $k$ values are more suitable for detecting and characterizing slight deviations in BH hair properties.

\begin{table*}[]
\centering
\begin{tabular}{p{1.2cm}p{1.2cm}p{1.5cm}p{1.5cm}p{1.5cm}p{1.5cm}p{1.5cm}p{1.5cm}p{1.5cm}p{1.5cm}p{1.5cm}}
\hline\hline
\rule{0pt}{12pt}
& & \multicolumn{6}{c}{Lensing Coefficients}&  &\\
\cline{3-8}
\rule{0pt}{12pt}
$a$&$Q_m$ &\multicolumn{3}{l}{$\bar{a}$} &  \multicolumn{3}{l}{$\bar{b}$} &\multicolumn{1}{l}{$u_m/R_s$} &\multicolumn{1}{l}{$u_m/R_s$}&\multicolumn{1}{l}{$u_m/R_s$}\\
\cline{3-8}
\rule{0pt}{11pt} 
  & &$k=1$ &  $k=1.5$&$k=2$ &$k=1$&$k=1.5$&$k=2$&$k=1$&$k=1.5$&$k=2$\\
\hline
-0.2&	0	&0.8796 &0.8796 &0.8796 &-0.2810 &-0.2810 &-0.2810&	2.9788 &	2.9788 &	2.9788\\ 
&0.3&0.9048&0.8911	&0.8833 &	-0.2570 &-0.2824 & -0.2838 &2.8331 &2.9543 &2.9746 \\
&0.4&0.9354 &0.9098 &0.8917 &-0.2382 &-0.2875 &-0.2907 &2.7043 &2.9186&2.9653 \\
&0.5&1.0152&	0.9533&0.9123 &-0.2388 &	-0.3123&-0.3106 &2.5054 &2.8533 &2.9446\\
\hline
0&0&	1.0000& 	1.0000 & 	1.0000&	-0.4002 &	-0.4002  &	-0.4002&	2.5981& 	2.5981& 	2.5981 \\
&	0.3	&1.0518& 	1.0220&1.0074& 	-0.3965& 	-0.4124 &-0.4048&	2.4294& 	2.5658&2.5918 \\
&	0.4	&1.1232& 	1.0609&1.0252& 	-0.4136& 	-0.4408&-0.4280& 	2.2730& 	2.5176 &2.5777\\
&	0.5	&1.4142& 	1.1714&1.0733& 	-0.7332& 	-0.5677&-0.4920& 	2.0000& 	2.4231&2.5452 \\
\hline
0.2&	0&	1.2209& 	1.2209& 	1.2209& 	-0.7700& 	-0.7700& 	-0.7700& 	2.1686&2.1686&2.1686 \\
&	0.3&	1.3717& 	1.2802&1.2423& 	-0.9488& 	-0.8508&-0.8051& 	1.9534& 	2.1200&2.1573 \\
&	0.4&	1.6922& 	1.4096&1.2989& 	-1.4988& 	-1.0636&-0.9060& 	1.7232& 	2.0413&2.1310 \\
	&0.5	&	-&	-&	-&	-		&	-	&-&-&-&-\\
\hline\hline
\end{tabular}
\caption{
The values corresponding to the lensing coefficients under specific spin parameters and hair parameters. When $a=0$ and $Q_m=0$, the BH degenerates to a Schwarzschild BH; when $a=0$ and $Q_m\neq0$, the BH becomes a Reissner-Nordström BH; when $a\neq0$ and $Q_m=0$, the BH is a Kerr BH. The parameter $k=1$ corresponds to the Kerr-Newman BH, while $k=1.5$ and $k=2$ correspond to rotating short-haired BH solutions under different parameter configurations.
}
\label{table1}
\end{table*}

\section{\label{sec:level4}Observational Effects of Strong Gravitational Lensing}

\subsection{\label{sec:level4.1}Characteristic Observation Values and Time Delays in Strong Lensing Effects}

In gravitational lensing, if the positions of the lens and the light source are known, the specific positions of the images can be calculated using the lens equation. The lens equation is given in \cite{Virbhadra:1999nm}, and later approximated using the small-angle approximation \cite{Bozza:2008ev,Bozza:2001xd}
\begin{equation}
\beta = \theta - \frac{D_{LS}}{D_{OS}} \Delta \alpha_n.
\label{41}
\end{equation}
Here, \(\beta\) represents the angle between the source and the lens axis, \(\theta\) represents the angle between the image and the lens axis, \(D_{LS}\) is the distance from the source to the lens, and \(D_{OS}\) is the distance from the source to the observer. The relationship among the source, lens (black hole), and observer is given by \(D_{OS} = D_{OL} + D_{LS}\). It is worth noting that, since the rotating short-hair black hole is asymptotically flat, for convenience of analysis, the source, lens, and observer can be set on the same line. In the above equation, \(\Delta \alpha_n = \alpha(\theta) - 2n\pi\) represents the remaining deflection angle after the light ray has looped around the black hole \(n\) times.

To approximate the deflection \(\Delta \alpha_n\), we use the analysis from \cite{Bozza:2002af}. For the \(n\)-th image, the relationship is given by
\begin{equation}
 \theta_n = \theta_n^0 + \Delta \theta_n,
\label{42}
\end{equation}
where
\begin{equation}
\theta_n^0 = \frac{u_m (1 + e_n)}{D_{OL}} ,
\label{43}
\end{equation}
\begin{equation}
e_n = \exp\left(\frac{\bar{b} - 2n\pi}{\bar{a}}\right)  ,
\label{44}
\end{equation}
\begin{equation}
 \Delta \theta_n = \frac{D_{OS}}{D_{LS}} \frac{u_m e_n}{\bar{a} D_{OL}} (\beta - \theta_n^0)  .
\label{45}
\end{equation}
Here, \(\theta_n^0\) corresponds to the angle \(\theta\) when \(\alpha = 2n\pi\). By combining these relationships, the approximate position of the \(n\)-th image can be obtained as \cite{Bozza:2002zj}
\begin{equation}
 \theta_n = \theta_n^0 + \frac{u_m e_n (\beta - \theta_n^0) D_{OS}}{\bar{a} D_{LS} D_{OL}} .
\label{46}
\end{equation}
Note that the relativistic image obtained from equation (\ref{46}) is only for the image on one side. For the image on the other side, we can obtain it in the same form by using $-\beta$. It can be seen from equation (\ref{46}) that the latter term is just a correction to \(\theta_n^0\).

Of course, in a lensing system, apart from the corresponding angular positions, the magnification is also an observable physical quantity of great significance. The magnification of the \(n\) -th image can be expressed as \cite{Bozza:2002zj,Virbhadra:1998dy,Virbhadra:2007kw}
\begin{equation}
\mu_n = \left( \frac{\beta}{\theta} \frac{d\beta}{d\theta} \right)^{-1} \Bigg|_{\theta_n^0} = \frac{u_m^2 (1 + e_n) D_{OS}}{\bar{a} \beta D_{LS} D_{OL}^2} e_n.
\label{49}
\end{equation}
It is evident from the above expression that due to the presence of the \(e_n\) term, the magnification of the image decreases exponentially with increasing \(n\). Additionally, when \(\beta\) is very small and approaches zero, we can obtain relatively bright images. In other words, images are easier to observe when \(\beta \to 0\). Clearly, the magnification is largest when \(n = 1\), meaning the image is the brightest at this point. Therefore, if we resolve the first-order image \(\theta_1\) (the dominant image) and represent the other unresolved images as \(\theta_\infty\), we can obtain several interesting observables \cite{Bozza:2002zj}
\begin{equation}
\theta_\infty = \frac{u_m}{D_{OL}} ,
\label{50}
\end{equation}

\begin{equation}
 S = \theta_1 - \theta_\infty = \theta_\infty \exp\left(\frac{\bar{b} - 2\pi}{\bar{a}}\right) ,
\label{51}
\end{equation}

\begin{equation}
 r_{\text{mag}} = \frac{\mu_1}{\sum_{n=2}^\infty \mu_n} \approx \frac{5\pi}{\bar{a} \ln(10)} .
\label{52}
\end{equation}
Here, \(\theta_\infty\) represents the angular position of the unresolved bundled images, \(S\) denotes the angular separation between the first resolved image (the outermost image) and the bundled images, and \(r_{\text{mag}}\) is the brightness ratio between the outermost image and the remaining bundled images. 
A detailed discussion of these observables is presented in Section \ref{sec:level4.2}.

In addition to the observational effects mentioned above, time delay is also an important physical quantity. In the study of gravitational lensing effects, when light passes by a BH, its path bends due to gravity, forming multiple images. These images, because of different light paths, reach the observer at different times, which is known as time delay. This phenomenon can be used to determine the geometric scale and mass of the lensing system in observations and to estimate the Hubble constant in a cosmological context, such as \cite{Birrer:2022chj,Treu:2022aqp,Grillo:2018ume}.Here, we use the method proposed by Bozza and Mancini for calculating time delay in the strong field limit \cite{Bozza:2003cp}. For a lensing system aligned on a straight line (\(\beta = 0\)) and two relativistic images on the same side of the lens (the \(p\)-th and \(q\)-th images), the time delay formula is given by \cite{Bozza:2003cp}
\begin{equation}
\Delta T_{p,q} = 2\pi (p - q) \frac{\tilde{a}}{\bar{a}} + 2 \sqrt{\frac{A_m u_m}{B_m}} \sqrt{u_m} \left( e^{-\frac{\bar{b} - 2p\pi}{2\bar{a}}} - e^{-\frac{\bar{b} - 2q\pi}{2\bar{a}}} \right) .
\label{53}
\end{equation}
Analyzing the above equation, the first term represents the geometric time delay, which mainly depends on the number of loops the light makes around the black hole. The second term represents the time dilation effect of the light in the gravitational field. It is evident that the time delay is primarily determined by the difference in the number of loops the light makes around the BH, so the dominant term is the first term. Thus, the above equation can be approximated as
\begin{equation}
\Delta T_{p,q} \approx 2\pi (p - q) \frac{\tilde{a}}{\bar{a}} = 2\pi (p - q) u_m = 2\pi (p - q) \theta_\infty D_{OL} .
\label{54}
\end{equation}

In the above discussion, we put forward several interesting observables, such as three key observable quantities and time delays. Based on these expressions, we can evaluate the observational results of the lensing effect in the real cosmic environment. In Section \ref{sec:level4.2}, we will take the known supermassive BHs M87* and Sgr A* in the real universe as our observation subjects. By using the relevant parameters of these actual BHs, we will assess the observational values of the lensing effect in the context of the rotating short-haired BH.

\subsection{\label{sec:level4.2}Evaluating the Observability of Supermassive Black Holes}

In this subsection, we will consider the rotating short-haired BH as a candidate for the supermassive BHs M87* and Sgr A* in the universe, and study its corresponding observables.  
For the supermassive BH M87* in the universe, the latest astronomical observational data shows that the mass of M87* is \((6.5\pm0.7)\times10^{9}M_{\odot}\), and its distance from the Earth is \((16.8\pm0.8)\text{Mpc}\) \cite{EventHorizonTelescope:2019ggy}. For the supermassive BH Sgr A*, its mass is \(3.98\times10^{6}M_{\odot}\), and the distance is \(7.97\text{kpc}\) \cite{Chen:2019tdb}. 
If the rotating short-haired BH is regarded as a candidate for these two supermassive BHs, relevant information ( \(D_{OL}\) and \(M\)) can be indirectly obtained. With this information, we can evaluate the observables calculated in the previous subsection.

\begin{figure*}[]
\includegraphics[width=0.9 \textwidth]{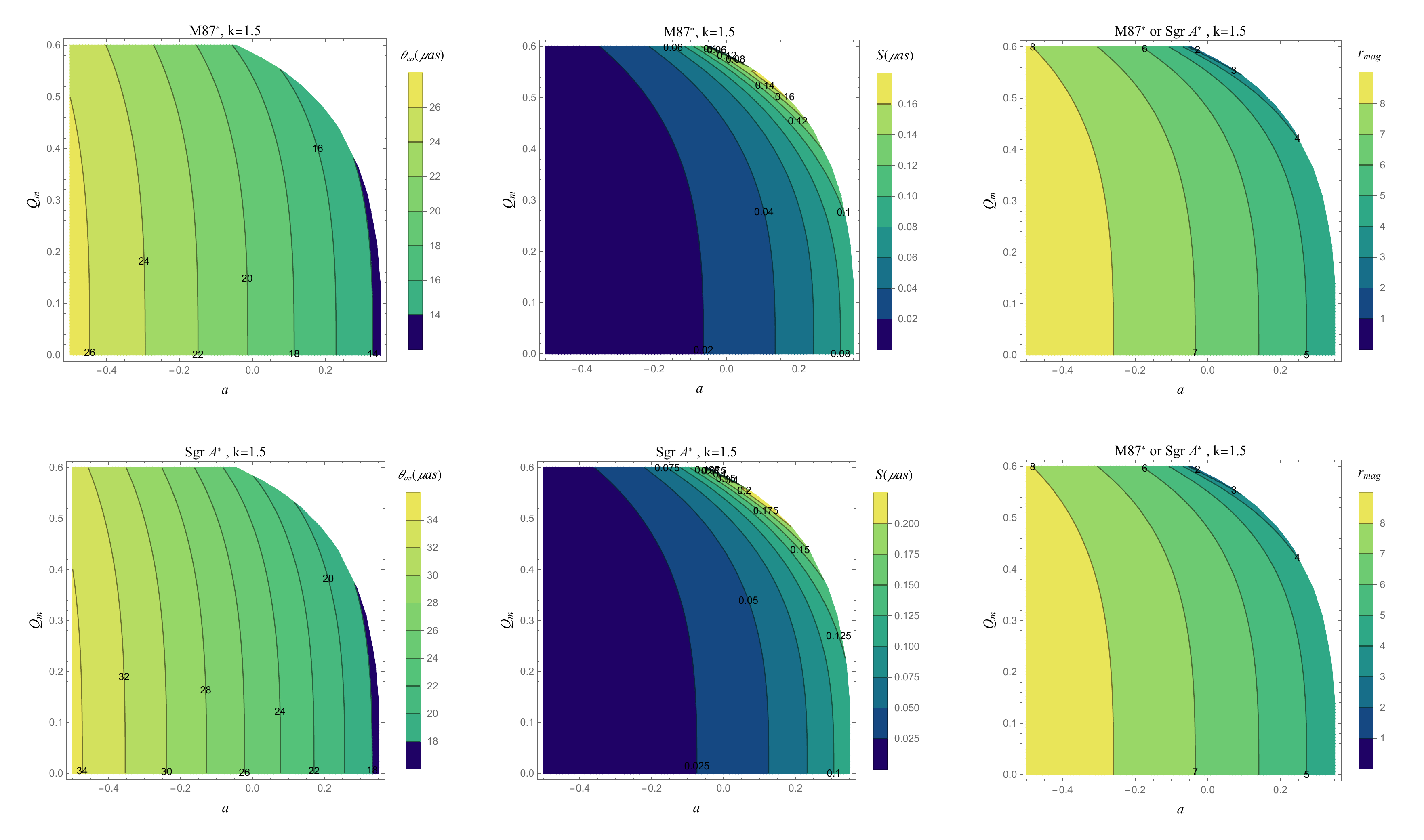}
\caption{
Variations of the Three Observables in M87* and Sgr A* with the Spin Parameter and Hair Parameter. The first row corresponds to M87*, and the second row corresponds to Sgr A*. Here, k = 1.5 represents the rotating short-haired BH.}
\label{l}
\end{figure*}

For the three interesting observables, when regarding the rotating short-haired BH as a candidate for M87* and Sgr A*, we carried out numerical calculations and presented the changes of these observables.
As illustrated in Figure \ref{l}, we selected the case of a rotating short-haired BH corresponding to parameter $k = 1.5$, and calculated the angular position of relativistic images \(\theta_{\infty}\), angular separation $S$, and image magnification ratio  $r_{\text{mag}}$ under different hair parameters $Q_m$.
The results show that, whether for Sgr A* or M87*, the variation trends of the observables are identical. 
Specifically, the angular position \(\theta_{\infty}\) of the relativistic image decreases as the spin parameter \(a\) and the hair parameter \(Q_m\) increase. The angular separation \(S\) increases with the increase of \(a\) and \(Q_m\). The image magnification ratio \(r_{\mathrm{mag}}\) gradually decreases as \(a\) increases. In particular, as \(Q_m\) increases, the variation trend becomes more significant (this is reflected by the density of the contour lines in the figure).
By combining the data in Table \ref{table2}, it can be seen that for M87*, the range of influence of the hair parameter \(Q_m\) on the angular position of the relativistic image is \(22.7113\ \mu as>\theta_{\infty}(M87*)>13.9162\ \mu as\). This range lies within the measurement range of \((42\pm3)\ \mu as\) when the EHT observes the diameter of M87* \cite{EventHorizonTelescope:2019dse}.
Regarding Sgr A*, the range of influence of the hair parameter is \(29.3132\ \mu as>\theta_{\infty}(Sgr A*)>17.9614\ \mu as\), and this range coincides with the measured value of \((51.8\pm2.3)\ \mu as\) for the diameter of the Sgr A* shadow observed by the EHT \cite{EventHorizonTelescope:2022wkp}.
Obviously, these observed values are all at the \(\mu as\) level, matching the observation accuracy of the existing EHT. Regrettably, though, the resolution of the current equipment is insufficient to detect these differences.

As can be seen from Table \ref{table3}, under the same spin parameter conditions, for M87*, the deviations in observables between the rotating short-haired BH and the Kerr BH (\(Q_m = 0\)) are \(\delta\theta_{\infty}\approx3\ \mu as\), \(\delta S\approx0.1272\ \mu as\), \(\delta r_{mag}\approx0.8980\); and the deviations between the rotating short-haired BH and the Kerr-Newman black hole (\(k = 1\)) are \(\delta\theta_{\infty}\approx2.4249\ \mu as\), \(\delta S\approx0.0535\ \mu as\), \(\delta S_{mag}\approx1.1001\).
For Sgr A*, the deviations between the rotating short-haired BH and the Kerr BH are \(\delta\theta_{\infty}\approx3.3786\ \mu as\), \(\delta S\approx0.1642\ \mu as\), \(\delta r_{mag}\approx0.8980\); and the deviations between the rotating short-haired BH and the Kerr-Newman BH are \(\delta\theta_{\infty}\approx3.1299\ \mu as\), \(\delta S\approx0.0690\ \mu as\), \(\delta S_{mag}\approx1.1001\).
Although these deviations are all at the \(\mu as\) level (especially the angular position), the angular resolution of the current EHT is approximately \(20\ \mu as\), so it is impossible to distinguish these small differences \cite{EventHorizonTelescope:2019ths}.
Therefore, to distinguish between the rotating short-haired BH and the Kerr BH or the Kerr-Newman BH may depend on the next-generation EHT. Once these differences can be distinguished in the future, it will provide us with further information about the short-hair parameter \(Q_m\), and also offer an opportunity to test the no-hair theorem.

\begin{figure*}[]
\includegraphics[width=0.9 \textwidth]{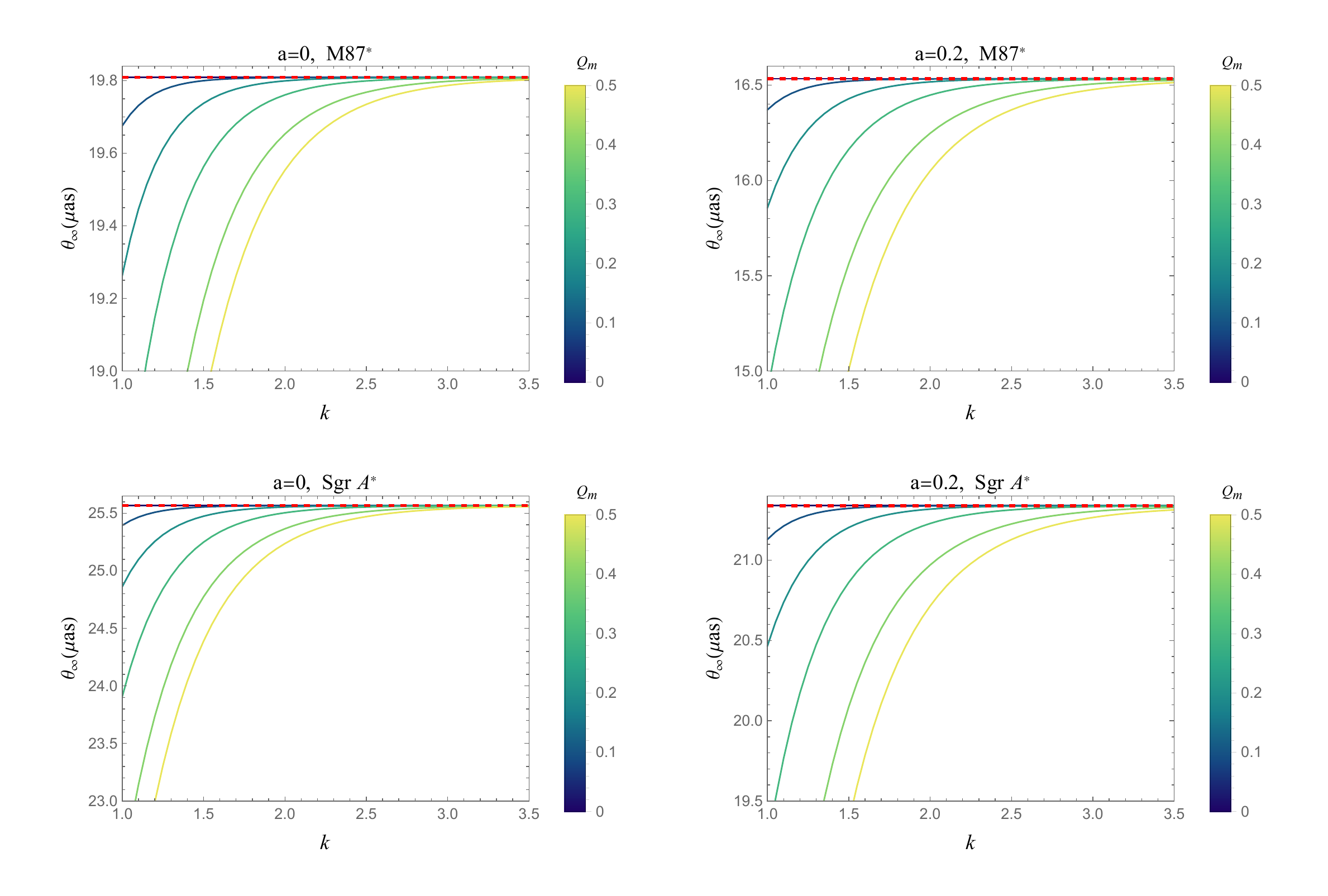}
\caption{
When considering M87* and Sgr A* as a rotating short-haired BH, the variation of angular position \(\theta_{\infty}\) with parameter $k$ under different hair parameters. The top row is for M87*, and the bottom row is for Sgr A*. The left column corresponds to the case where $a=0$, and the right column corresponds to the case where $a=0.2$. The red dotted line in the figure represents the case of a Kerr BH.}
\label{a1}
\end{figure*}

In addition, from Figure \ref{a1} and Table \ref{tablea}, it can be observed that as the parameter $k$ increases, all corresponding observables gradually approach the case of the standard Kerr BH (the red dashed line in Figure {a1} and the values shown in the header of Table \ref{tablea} represent the Kerr BH case with $Q_m=0$). This result indicates that larger $k$ values actually suppress the influence of the hair parameter, causing the differences in observational characteristics between rotating short-haired BH  and Kerr BH to gradually diminish to an indistinguishable degree. Based on this consideration, we chose to fix the parameter $k=1.5$ in the above analysis, a choice that both maintains the rationality of the physical model and adequately demonstrates the characteristic influence of the hair parameter on observables. This allows us to effectively quantify and evaluate the contributions of hair parameters and spin to various observables.

\begin{table*}[]
\centering
\begin{tabular}{p{1cm}p{1cm}p{1.4cm}p{1.4cm}p{1.4cm}p{1.4cm}p{1.4cm}p{1.4cm}p{1.4cm}p{1.5cm}p{1.4cm}p{1.4cm}}
\hline\hline
\rule{0pt}{12pt}
& &\multicolumn{4}{l}{$M87^*$} & \multicolumn{4}{l}{$Sgr A^*$} & \multicolumn{2}{l}{$M87^* or Sgr A^*$}\\
\cline{3-12}
\rule{0pt}{12pt}
& &\multicolumn{2}{l}{$k=1$}&\multicolumn{2}{l}{$k=1.5$}&\multicolumn{2}{l}{$k=1$}&\multicolumn{2}{l}{$k=1.5$}&\multicolumn{1}{l}{$k=1$}&\multicolumn{1}{l}{$k=1.5$}
\\
\cline{3-12}
\rule{0pt}{12pt}
$a$& $Q_m$& $\theta_\infty(\mu as)$& $S(\mu as)$ & $\theta_\infty(\mu as)$ &  $S(\mu as)$& $\theta_\infty(\mu as)$ & $S(\mu as)$& $\theta_\infty(\mu as)$ & $S(\mu as)$ & $r_{mag}$ & $r_{mag}$  
\\
\hline
\rule{0pt}{11pt} 
-0.2&	0&	22.7113 &	0.0130& 	22.7113 &	0.0130& 	29.3132& 	0.0168& 	29.3132& 	0.0168& 	7.7554& 	7.7554 \\
&	0.3&	21.6003& 	0.0157& 	22.5243& 	0.0142& 	27.8792& 	0.0202& 	29.0718& 	0.0183& 	7.5398& 	7.6559\\ 
&	0.4&	20.6186& 	0.0193& 	22.2525& 	0.0163& 	26.6121& 	0.0250& 	28.7210& 	0.0210& 	7.2930& 	7.4980\\ 
&	0.5&	19.1019& 	0.0310& 	21.7547& 	0.0215& 	24.6545& 	0.0400& 	28.0784& 	0.0278& 	6.7200& 	7.1564 
\\
\hline
0&	0&	19.8085 &	0.0248& 	19.8085& 	0.0248& 	25.5668& 	0.0320& 	25.5668& 	0.0320& 	6.8219& 	6.8219\\ 
&	0.3&	18.5222& 	0.0323& 	19.5629& 	0.0279& 	23.9064& 	0.0417& 	25.2495& 	0.0361& 	6.4858& 	6.6750\\ 
&	0.4&	17.3301& 	0.0446& 	19.1951& 	0.0339& 	22.3677& 	0.0576& 	24.7749& 	0.0438& 	6.0738& 	6.4300 \\
&	0.5&	15.2487& 	0.1068& 	18.4742& 	0.0533& 	19.6813& 	0.1378& 	23.8444& 	0.0688& 	4.8238& 	5.9239\\
\hline
0.2&	0&	16.5339& 	0.0512& 	16.5339& 	0.0512& 	21.3400& 	0.0661& 	21.3400& 	0.0661& 	5.5874& 	5.5874 \\
&	0.3&	14.8931& 	0.0764& 	16.1635& 	0.0614& 	19.2224& 	0.0986& 	20.8620& 	0.0793& 	4.9733& 	5.3286 \\
&	0.4&	13.1383& 	0.1322& 	15.5632& 	0.0848& 	16.9574& 	0.1707& 	20.0873& 	0.1095& 	4.0314& 	4.8396\\ 
&	0.5&-	&- &13.9162& 	0.1784&- &- &	17.9614& 	0.2303&- &	2.2650
\\
\hline\hline
\end{tabular}
\caption{Observed values corresponding to different hairy parameters \(Q_m\) when considering M87* and Sgr A* as the rotating short-haired BH. Here, \(k = 1\) corresponds to the Kerr-Newman BH, and \(k = 1.5\) corresponds to the rotating short-haired BH.}
\label{table2}
\end{table*}

\begin{table*}[]
\centering
\begin{tabular}{p{1cm}p{1cm}p{1.4cm}p{1.4cm}p{1.4cm}p{1.4cm}p{1.4cm}p{1.4cm}p{1.4cm}p{1.5cm}p{1.4cm}p{1.4cm}}
\hline\hline
\rule{0pt}{12pt}
& &\multicolumn{4}{l}{$M87^*$} & \multicolumn{4}{l}{$Sgr A^*$} & \multicolumn{2}{l}{$M87^* or Sgr A^*$}\\
\cline{3-12}
\rule{0pt}{12pt}
& &\multicolumn{2}{l}{$Q_m=0.1$}&\multicolumn{2}{l}{$Q_m=0.3$}&\multicolumn{2}{l}{$Q_m=0.1$}&\multicolumn{2}{l}{$Q_m=0.3$}&\multicolumn{1}{l}{$Q_m=0.1$}&\multicolumn{1}{l}{$Q_m=0.3$}
\\
\cline{3-12}
\rule{0pt}{12pt}
$a$& $k$& $\theta_\infty(\mu as)$& $S(\mu as)$ & $\theta_\infty(\mu as)$ &  $S(\mu as)$& $\theta_\infty(\mu as)$ & $S(\mu as)$& $\theta_\infty(\mu as)$ & $S(\mu as)$ & $r_{mag}$ & $r_{mag}$  
\\
\hline
\rule{0pt}{11pt} 
0&1&	19.6756&	0.0254&	18.5222&	0.0323&	25.3950&	0.0328&	23.9064&	0.0417&	6.7909&	6.4857 \\
&1.5&19.7998 &0.0249 &19.5629&	0.0279&	25.5554 &0.0322 &25.2495&0.0361 &6.8168 &6.6750 \\
&2&19.8081 &0.0248&	19.7606&	0.0258 &	25.5660&	0.0320&	25.5047&	0.0333&	6.8213&	6.7716\\ 
&2.5&19.8086&	0.0248&	19.7991&	0.0251&	25.5667&	0.0320&	25.5545&	0.0324&	6.8218&	6.8065 \\
&3&19.8087&	0.0248 &19.8068 &0.0249 &25.5668&	0.0320&25.5643&	0.0321&	6.8219 &	6.8175\\
\hline
\rule{0pt}{11pt} 
0.2&1&16.3708 &	0.0531&	14.8931 &0.0764 &21.1295&	0.0685&	19.2224&	0.0986&	5.5342&	4.9733 \\
&1.5&16.5209&	0.0515&	16.1635&	0.0614&	21.3233&	0.0665&	20.8620&	0.0793&	5.5788&	5.3286 \\
&2&16.5328&0.0513 &	16.4479&	0.0547&	21.3387&	0.0662& 	21.2291& 	0.0706& 	5.5862& 	5.4914 \\
&2.5&	16.5338 &	0.0512 &	16.5135& 	0.0524& 	21.3399& 	0.0661& 	21.3137& 	0.0676& 	5.5872& 	5.5541 \\
&3&	16.5339& 	0.0512& 	16.5290 &	0.0516& 	21.3400& 	0.0661& 	21.3337 &	0.0666& 	5.5874& 	5.5764
\\
\hline\hline
\end{tabular}
\caption{
The observable values corresponding to different parameters $k$ under fixed hair parameters $Q_m=0.1$ or $Q_m=0.3$ when considering M87* and Sgr A* as a rotating short-haired BH. For comparison reference, the standard Kerr BH $(Q_m=0)$ at spin parameter $a=0$ has corresponding observable values of $\theta_\infty=19.8087\,\mu as$, $S=0.0248\,\mu as$, $r=6.8219 \,\mu as$ (for M87*), $\theta_\infty=25.5668\,\mu as$, $S=0.0320\,\mu as$, $r=6.8219 \,\mu as$ (for Sgr A*); at spin parameter $a=0.2$, its observable parameters are $\theta_\infty=16.5339\,\mu as$, $S=0.0512\,\mu as$, $r=6$ (for M87*), $\theta_\infty=21.3400\,\mu as$, $S=0.0661\,\mu as$, $r=5.5874$ (for Sgr A*).
}
\label{tablea}
\end{table*}

\begin{table*}[]
\centering
\begin{tabular}{p{1cm}p{1cm}p{1.4cm}p{1.4cm}p{1.4cm}p{1.4cm}p{1.4cm}p{1.4cm}p{1.4cm}p{1.5cm}p{1.4cm}p{1.4cm}}
\hline\hline
\rule{0pt}{12pt}
& &\multicolumn{4}{l}{$M87^*$} & \multicolumn{4}{l}{$Sgr A^*$} & \multicolumn{2}{l}{$M87^* or Sgr A^*$}\\
\cline{3-12}
\rule{0pt}{12pt}
& & \multicolumn{2}{l}{$hair- Kerr$}&\multicolumn{2}{l}{$hair- KN$}&\multicolumn{2}{l}{$hair- Kerr$}&\multicolumn{2}{l}{$hair- KN$}&\multicolumn{1}{l}{$hair- Kerr$}&\multicolumn{1}{l}{$hair- KN$}
\\
\cline{3-12}
\rule{0pt}{12pt}
$a$& $\delta Q_m$& $\delta\theta_\infty(\mu as)$& $\delta S(\mu as)$ & $\delta\theta_\infty(\mu as)$ &  $\delta S(\mu as)$& $\delta\theta_\infty(\mu as)$ & $\delta S(\mu as)$& $\delta\theta_\infty(\mu as)$ & $\delta S(\mu as)$ & $\delta r_{mag}$ & $\delta r_{mag}$  
\\
\hline
\rule{0pt}{11pt} 
-0.2& 0.3&	0.1870& 	0.0012& 	0.924&	0.0015& 	0.2414& 	0.0015& 	1.1926& 	0.0019& 	0.0995& 	0.1161\\ 
&	0.4&	0.4588& 	0.0032& 	1.6339&	0.0030& 	0.5922& 	0.0041& 	2.1089& 	0.0040& 	0.2574& 	0.2050 \\
&	0.5&	0.9566& 	0.0085& 	2.6528&	0.0095& 	1.2348& 	0.0109& 	3.4239& 	0.0122& 	0.5990& 	0.4364 
\\
\hline
0&0.3&	0.2456& 	0.0031& 	1.0407&	0.0044& 	0.3173& 	0.0041& 	1.3431& 	0.0056& 	0.1469& 	0.1892\\ 
&	0.4&	0.6134& 	0.0092& 	1.865&	0.0107& 	0.7919& 	0.0118& 	2.4072& 	0.0138& 	0.3919& 	0.3562\\ 
&	0.5&	1.3343& 	0.0285& 	3.2255&	0.0535& 	1.7224& 	0.0368& 	4.1631& 	0.0690& 	0.8980& 	1.1001 
\\
\hline
0.2&	0.3&	0.3704& 	0.0102& 	1.2704&	0.0150& 	0.4780& 	0.0132& 	1.6396& 	0.0193& 	0.2588& 	0.3553\\ 
&	0.4&	0.9707& 	0.0336& 	2.4249&	0.0474& 	1.2527& 	0.0434& 	3.1299& 	0.0612& 	0.7478& 	0.8082\\ 
&0.5&	2.6177& 	0.1272 &- &	-	&	3.3786& 	0.1642& -&- &	3.3224&	-	\\	

\hline\hline
\end{tabular}
\caption{When considering M87* and Sgr A* as the rotating short-haired BH, the deviations in the observed values of the rotating short-haired BH from those in the cases of Kerr BH and Kerr-Newman BH. Here, \(\delta(X)=X_{\mathrm{hair}}^{k=1.5} - X_{\mathrm{Kerr}}\) or \(\delta(X)=X_{\mathrm{hair}}^{k=1.5} - X_{\mathrm{KN}}\).}
\label{table3}
\end{table*}

\begin{table*}[]
\centering
\begin{tabular}{p{1.2cm}p{1.2cm}p{2cm}p{2cm}p{2cm}p{2cm}p{2cm}p{2cm}}
\hline\hline
\rule{0pt}{12pt}
& &\multicolumn{3}{l}{$M87^*$} & \multicolumn{3}{l}{$Sgr A^*$}\\
\cline{3-8}
\rule{0pt}{12pt}
& & \multicolumn{1}{l}{$k=1$}& \multicolumn{1}{l}{$k=1.5$}& \multicolumn{1}{l}{$k=2$}&\multicolumn{1}{l}{$k=1$}&\multicolumn{1}{l}{$k=1.5$}&\multicolumn{1}{l}{$k=2$}
\\
\cline{3-8}
\rule{0pt}{12pt}
$a$& $Q_m$& $\Delta T_{21}(h)$& $\Delta T_{21}(h)$& $\Delta T_{21}(h)$ & $\Delta T_{21}(min)$ &  $\Delta T_{21}(min)$&  $\Delta T_{21}(min)$
\\
\hline
\rule{0pt}{11pt} 
-0.2&	0&	332.1074& 	332.1074&332.1074& 	12.2011& 	12.2011&12.2011\\ 
&	0.3&	315.8604& 	329.3723&331.637& 	11.6042& 	12.1006&12.1838\\ 
&	0.4&	301.5053& 	325.3982&330.5983& 	11.0768& 	11.9546&12.1457\\ 
&	0.5&	279.3262& 	318.1182&328.2968& 	10.2620& 	11.6872&12.0611\\ 
\hline
0&	0&	289.6618& 	289.6618&289.6618& 	10.6417& 	10.6417&10.6417\\ 
&	0.3&	270.8501& 	286.0676&288.959& 	9.9506& 	10.5097&10.6159\\ 
&	0.4&	253.4180& 	280.6904&287.3847& 	9.3102& 	10.3121&10.5581\\ 
&	0.5&	222.9818& 	270.1475&283.7672& 	8.1920& 	9.9248&10.4252\\
\hline
0.2&	0&	241.7744& 	241.7744&241.7744& 	8.8824& 	8.8824&8.8824\\ 
&	0.3&	217.7821& 	236.3585&240.5179& 	8.0010& 	8.6834&8.8363\\ 
&	0.4&	192.1213& 	227.5810&237.5862& 	7.0582& 	8.3610&8.7286\\ 
&	0.5&-	&	203.4960&229.9303&- 	&	7.4761&8.4473\\
\hline\hline
\end{tabular}
\caption{
When considering M87* and Sgr A* as a rotating short-haired BH, the time delay \(\Delta T_{21}\) between two images on the same side.
}
\label{table4}
\end{table*}

\begin{figure*}[]
\includegraphics[width=0.9 \textwidth]{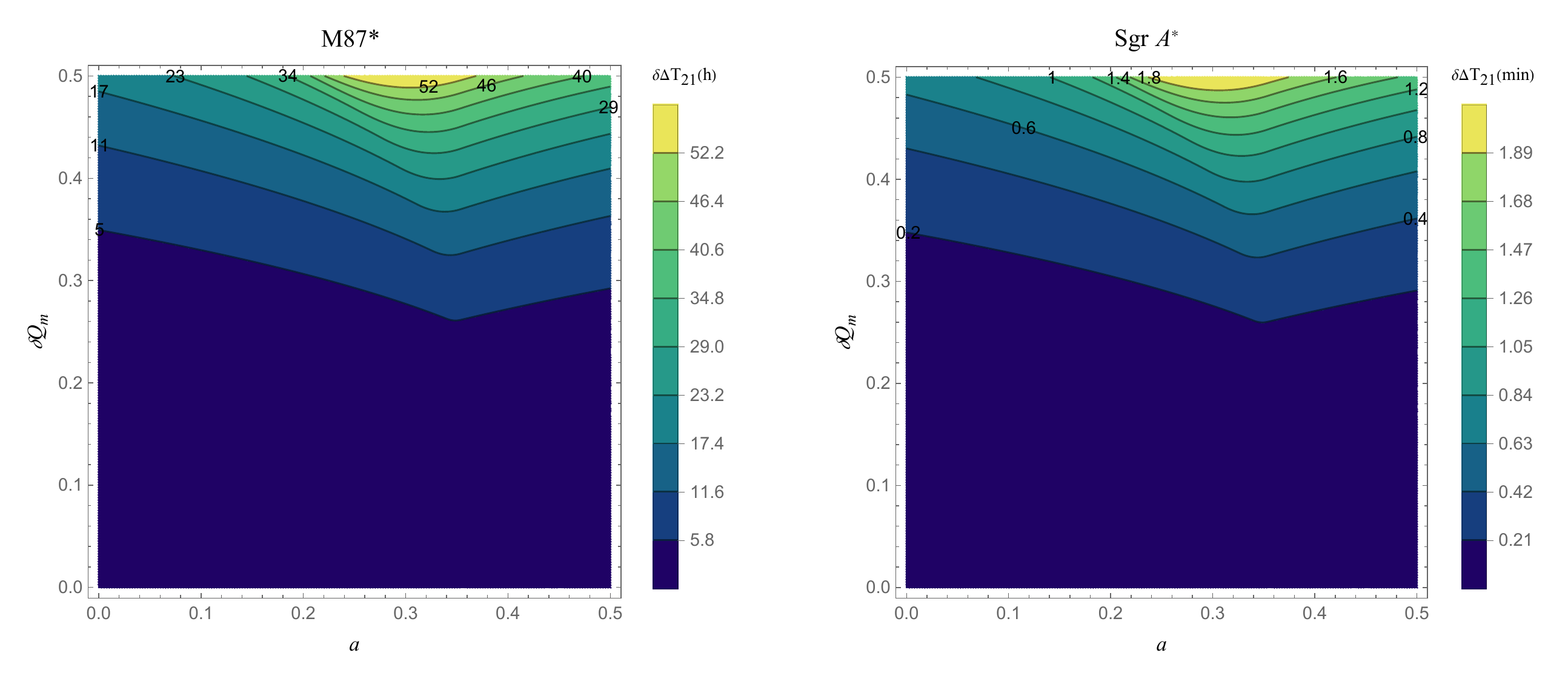}
\caption{
When considering M87* and Sgr A* as the rotating short-haired BH, the time  delay deviation \(\delta\Delta T_{21}\) of the same image between the rotating short-haired BH and the Kerr BH. Here, \(\delta(X)=X_{\mathrm{hair}}^{k=1.5} - X_{\mathrm{Kerr}}\).
}
\label{fig8}
\end{figure*}

\begin{figure*}[]
\includegraphics[width=0.9 \textwidth]{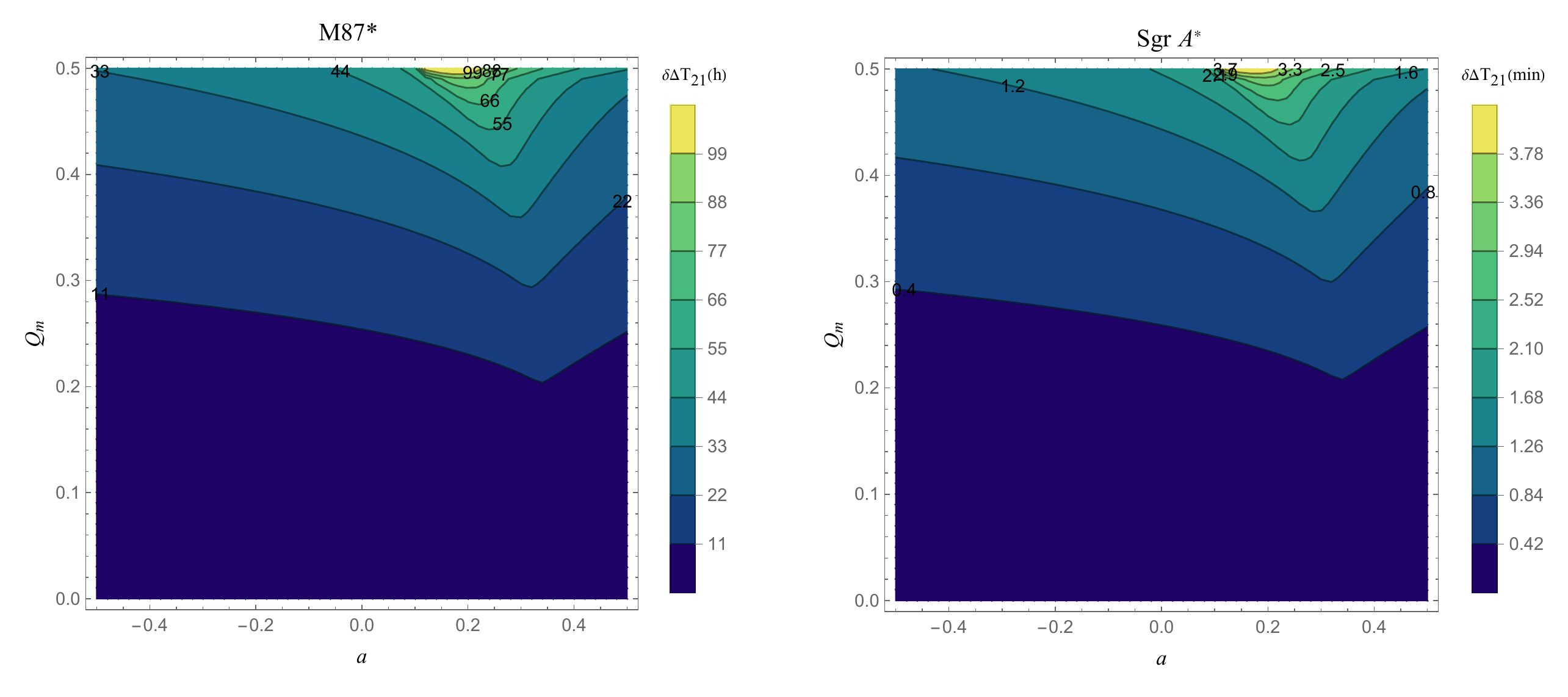}
\caption{
When M87* and Sgr A* are regarded as the rotating short-haired BH, the time delay deviation of the same image, \(\delta\Delta T_{21}\), between the rotating short-haired BH and the Kerr BH. Here, \(\delta(X)=X_{\mathrm{hairy}}^{k=1.5} - X_{\text{K-N}}\).
}
\label{fig9}
\end{figure*}

For the two relativistic images located on the same side of the BH, the time delay between the second relativistic image (\(p = 2\)) and the first relativistic image (\(q = 1\)) is presented in Table \ref{table4}. The deviations between the rotating short-haired BH ($k=1.5$), the Kerr BH, and the Kerr-Newman BH are shown in Table \ref{table4} as well as Figures \ref{fig8} and \ref{fig9}. Here, the time delay we consider is mainly determined by the optical path difference between relativistic images formed after light rays orbit the BH several times (see Figure \ref{f}. Obviously, near the divergence of the deflection angle, the light rays orbit more than \(2\pi\)). When considering M87* and SgrA* as a rotating short-haired BH, the time delay \(\mathrm{\Delta}T_{2,1}\) between the second and the first relativistic images on the same side is such that the time delay of the former can be up to several hundred hours, while that of the latter can reach tens of minutes. Evidently, the time delay of the former is sufficient for astronomical observations, which provides a necessary condition for exploring the properties of the short-haired BH (see Table \ref{table4}).
Under the same circumstances, the time  delay deviation amounts between the rotating short-haired BH on one hand, and the Kerr BH and the Kerr-Newman BH on the other hand: In the case of Sgr A* as the background, the time  delay deviation between the rotating short-haired BH and the Kerr BH can reach up to more than two minutes (see  Figure \ref{fig8}), while the time  delay deviation between the rotating short-haired BH and the Kerr-Newman BH can reach up to more than four minutes (see Figure \ref{fig9}). In the case of M87* as the background, the time  delay deviation between the rotating short-haired BH and the Kerr BH can reach up to more than fifty hours (see  Figure \ref{fig8}), and the time  delay deviation between the rotating short-haired BH and the Kerr-Newman BH can reach up to more than one hundred hours (see  Figure \ref{fig9}).
Overall, for Sgr A*, the time   delay deviation is in the range of several minutes, which is difficult to be detected in astronomy. For M87*, the time  delay can reach hundreds of hours. The time   delay deviation between it (when considering M87* as a rotating short-haired BH) and the Kerr BH can reach over fifty hours, and the deviation between it and the Kerr-Newman BH can reach over one hundred hours. These time durations are sufficient for astronomical observations. Just from the perspective of time  delay, if M87* is regarded as a candidate for a rotating short-haired BH, it is feasible to distinguish between traditional BHs and rotating short-haired BHs. However, the prerequisite for this is that the equipment used is capable of resolving the two relativistic images.
Obviously, current equipment fails to meet this requirement. But the next  generation EHT is expected to obtain such resolution. If these images can be distinguished in the near future, a lot of important physical information will be obtained, such as testing the no-hair theorem, exploring the properties of the  hair parameter \(Q_m\), and restricting the variation of parameters.

\section{\label{sec:level5}Constraints from the EHT}

\begin{figure*}[]
\includegraphics[width=0.9 \textwidth]{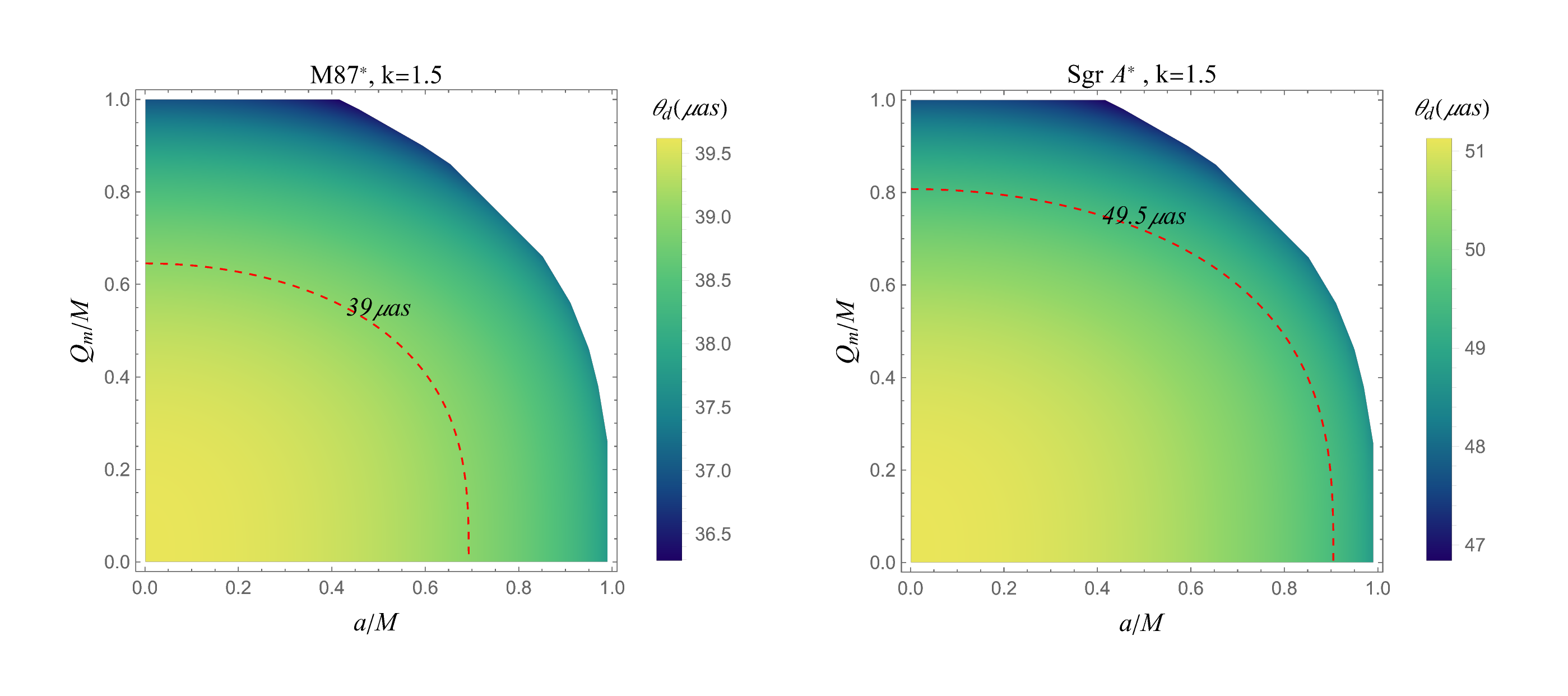}
\caption{
After taking the rotating short-haired BH model as a candidate model for M87* and Sgr A*, the parameter space ($Q_m/M, a/M$) is constrained using the EHT data. The left  hand figure corresponds to M87*, and the right  hand figure corresponds to Sgr A*. The red dotted line corresponds to the boundary of the $1\sigma$ confidence interval. Here, the inclination angle $\theta_o = \frac{\pi}{2}$. 
}
\label{fig10}
\end{figure*}
\begin{figure*}[]
\includegraphics[width=0.9 \textwidth]{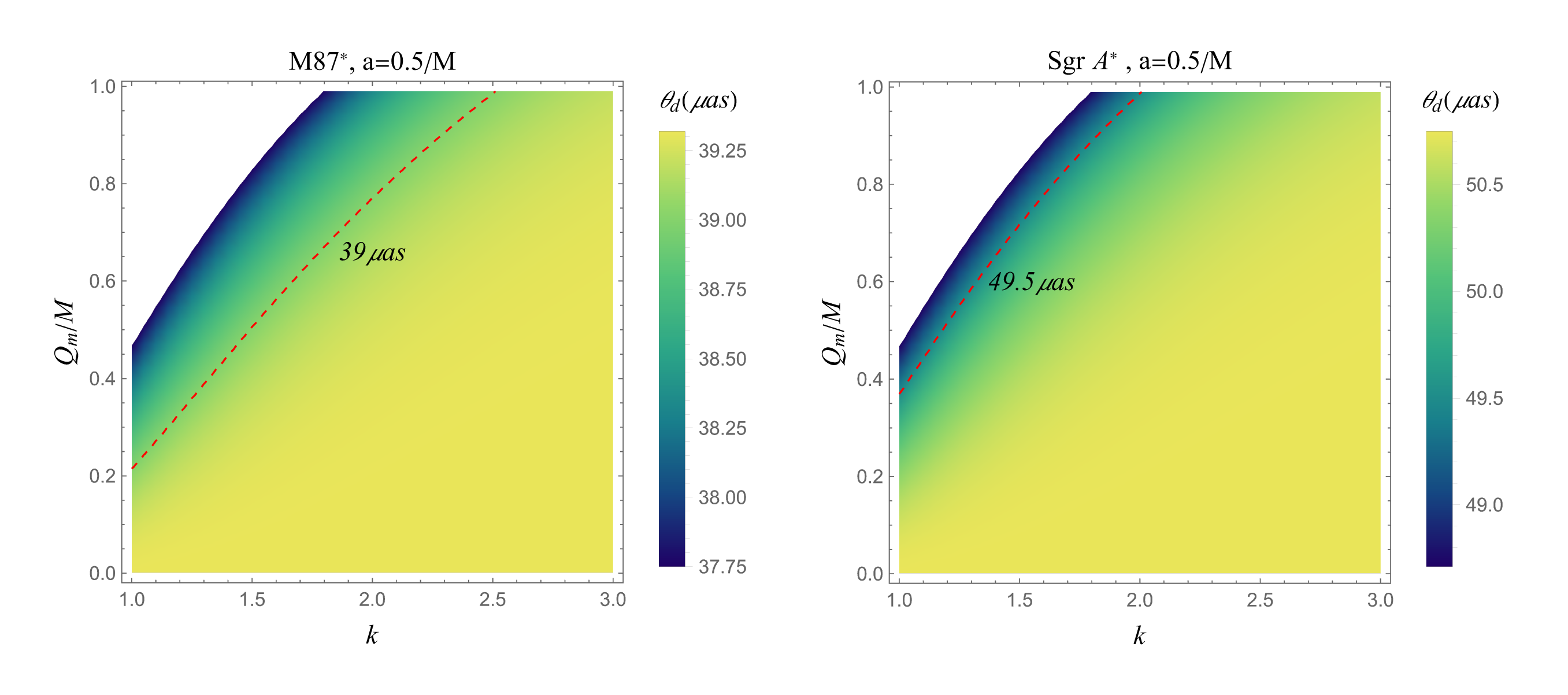}
\caption{
Parameter space constraints for rotating short-haired black hole models as candidate models for M87* and Sgr A*. Left figure: Parameter constraints for M87*; Right figure: Parameter constraints for Sgr A*. The red dashed line represents the boundary of the corresponding $1\sigma$ confidence interval. The analysis fixes the black hole spin parameter $a=0.5$ and the observation inclination angle $\theta_o=\frac{\pi}{2}$.
}
\label{fig12}
\end{figure*}

In the preceding section, we delved into the influence exerted by the hair parameter on the strong gravitational lensing phenomenon. The hair parameter acts by modifying the spacetime structure in the vicinity of the rotating short-haired BH, thereby inducing alterations in the propagation trajectories of light rays within the BH's gravitational field. As light rays approach the event horizon, the intense gravitational field of the black hole causes the light rays to undergo deflection, giving rise to the gravitational lensing effect. This may potentially result in the formation of multiple images or regions of enhanced luminosity, as elaborated in detail in the previous section. With further deflection of the light rays, a black hole shadow is ultimately formed.
The geometric properties of the shadow of a rotating short-haired BH are not only contingent upon the black hole's spin but are also significantly influenced by the hair parameter. Consequently, while having an impact on the propagation paths of light rays, the hair parameter exerts a far-reaching influence on both the shape and size of the black hole shadow. A comprehensive discussion regarding the effect of the hair parameter on the shape of the shadow surrounding a rotating short-haired BH can be found in the reference \cite{Tang:2022uwi}. Therefore, our emphasis does not lie in a detailed exploration of the effects on the shadow shapes. Instead, we are committed to systematically constraining the relevant parameter space of the rotating short-haired BH by leveraging the shadow data of supermassive BHs obtained from the observations of the EHT. 

In 2019, the EHT successfully captured the image of the black hole shadow of the supermassive BH M87* located at the center of the elliptical galaxy M87, which brought revolutionary progress to the observational and theoretical research of supermassive BHs. The EHT team measured the diameter of the black hole shadow of M87* to be \(\theta_d = 42\pm3\ \mu\mathrm{as}\) \cite{EventHorizonTelescope:2019ggy,EventHorizonTelescope:2019dse,EventHorizonTelescope:2019pgp}. This result not only provides us with precise observational evidence of the geometric structure of the BH but also offers evidence for verifying the validity of general relativity in extreme gravitational fields.
Subsequently, in 2022, the EHT team further enhanced the observational technology and data  processing capabilities and successfully measured the diameter of the black hole shadow of the supermassive BH Sgr A* located at the center of the Milky Way. The result is \(\theta_d = 51.8\pm2.3\ \mu\mathrm{as}\) \cite{EventHorizonTelescope:2022wkp}. These data play a crucial role in the detection of gravitational theories and parameter constraints, such as \cite{Jusufi:2021fek,Zahid:2023csk,Afrin:2022ztr,Raza:2023vkn,Sanchez:2024sdm,Zahid:2023csk}.

For axisymmetric black holes, the geometric shape of the black hole shadow is influenced by the spin parameter. To more accurately describe the shape of the shadow, an astronomical coordinate system \cite{Hioki:2009na} is usually introduced, which is defined as
\begin{equation}
X = \lim_{r_o\rightarrow\infty}\left(-r_o^2\sin\theta_o\frac{d\phi}{dr}\right),
\label{111}
\end{equation}

\begin{equation}
Y=\pm\lim_{r_o\rightarrow\infty}\left(r_o^2\frac{d\theta}{dr}\right),
\label{112}
\end{equation}
where \(r_o\) and \(\theta_o\) are the distance and inclination angle of the observer in the Boyer-Lindquist coordinate system, respectively. If considering the equatorial plane (\(\theta=\frac{\pi}{2}\)) and combining with the metric (\ref{2}) of the rotating short-haired BH, then formulas (\ref{111}) and (\ref{112}) can be rewritten as
\begin{equation}
X = \frac{4r\Delta-\Delta'r^2 + a^2}{a\Delta'},
\label{113}
\end{equation}
\begin{equation}
Y=\pm\frac{\sqrt{r^2(a^2 - \Delta)(16\Delta-r^2\Delta'^2 + 8r\Delta\Delta')}}{a^2\Delta'^2}.
\label{114}
\end{equation}
Here, the prime (') indicates the derivative with respect to \(r\), and \(\Delta\) is a function term associated with the rotating short-haired BH. The detailed derivation process of this part can be found in corresponding literature, such as \cite{Tang:2022uwi,Zahid:2023csk,Sarikulov:2022atq}, etc.

Once we obtain the corresponding expressions for celestial coordinates, it becomes straightforward to determine the shadow shape of a rotating short-haired black hole. This aspect has been extensively discussed in the literature \cite{Tang:2022uwi}; therefore, we shall not elaborate further on this topic here. The primary objective of this section is to constrain the hair parameter and spin parameter of the rotating short-haired black hole using the supermassive black hole shadow data observed by the EHT.

If the rotating short-haired BH model is considered as a candidate for M87* or SgrA*, then for a distant observer, the angular diameter $\theta_d$ of the shadow image around the rotating short-haired BH can be expressed as (e.g., \cite{Afrin:2021imp,Zahid:2023csk,Afrin:2022ztr,Kumar:2020owy}, etc.)
\begin{equation}
\theta_d=\frac{2R_a}{d}, \quad R_a^2=\frac{A}{\pi}.
\label{53}
\end{equation}
Where $d$ represents the distance to the observer, $R_a$ is the area shadow radius, and $A$ is the region within the shadow contour \cite{Sarikulov:2022atq,Kumar:2018ple}
\begin{equation}
A=\int_{r_-}^{r_+}{Y\frac{dX}{dr}dr},
\label{54}
\end{equation}
where $r_+$ and $r_-$ are the roots of $Y^2=0$ outside the event horizon radius, corresponding to the prograde and retrograde radii of stable circular orbits.

With these preparations, we can model the rotating short-haired BH model as M87* or Sgr A*, so as to constrain the parameter space \((Q_m/M, a/M)\) of the rotating short-haired BH by using the data observed by the EHT. As shown in the left  hand panel of Figure \ref{fig10}, considering the rotating short-haired BH model as a candidate for M87* and combining with the EHT observational data, we have constrained the parameter space of \(Q_m\) (hair parameter) and \(a\) (spin parameter) within the first confidence interval. The constrained region is enclosed by the red dashed line. The red curve in the figure, corresponding to an angular diameter \(\theta_d = 39\ \mu as\), represents the boundary of the first confidence interval. Within this confidence interval, the EHT data provides an effective constraint on the parameter space \((Q_m/M, a/M)\) of the rotating short-haired BH. Specifically, the constraint on the hair parameter is influenced by the spin parameter. As the \(a\) increases, the constrained range of the \(Q_m\) gradually shrinks.

Similarly, when applying the rotating short-haired BH model to Sgr A*, we can also utilize the EHT observational data of the Sgr A* shadow to constrain the parameter space of \(Q_m\) and \(a\) within the first confidence interval. The region enclosed by the red dashed line is the constrained interval. In the right figure of Figure \ref{fig10}, the red dashed line, corresponding to an angular diameter \(\theta_d = 49.5\ \mu as\), represents the boundary of the first confidence interval. As shown in the right figure, the constraints on the parameter space in Sgr A* are similar to those in M87*. Both the spin parameter \(a\) and the hair parameter \(Q_m\) are significantly constrained, and the constrained range of \(Q_m\) gradually narrows as \(a\) increases.

Furthermore, in order to explore the impact of parameter $k$ on the constraint of the hair parameter $Q_m$, we fixed the spin parameter $a = 0.5$ in Figure \ref{fig12} and systematically analyzed the changes in the constraint interval of $Q_m$ under different values of $k$. The results show that, regardless of whether it is based on the observational data constraints from M87* or Sgr A*, an increase in parameter $k$ leads to a remarkable weakening of the constraining ability on the hair parameter $Q_m$. (The red dashed line in the figure represents the $1\sigma$ confidence interval boundary.) Specifically, as the value of $k$ increases, the allowable interval for $Q_m$ gradually broadens, and the constraining power gradually diminishes. This conclusion is in high agreement with the analysis results regarding the influence of $k$ on observables such as the light deflection angle and the relativistic angular position of the image in the previous sections. It further validates that a degeneracy exists between the rotating short-haired BH and the Kerr BH in the region of large $k$ values.

Overall, whether using the shadow data of M87* or Sgr A* to constrain the parameter space \((Q_m/M, a/M)\) of the rotating short-haired BH model, our analysis results show that the parameter space of the rotating short-haired BH model is highly consistent with the range of the observational data from the EHT. Therefore, through a systematic analysis of the EHT observational results of M87* and Sgr A* within the first confidence interval, we cannot rule out the rotating short-haired BH model as a potential candidate for real  universe BHs.
In addition, although the current observational accuracy still has limitations, the EHT data has provided a certain detectability for the rotating short-haired BH model. With the gradual improvement of observational techniques, future data may help refine the parameter space of this model and provide new clues for distinguishing different BH models. This result supports the potential of the rotating short-haired BH model as a description of real universe BHs.

\section{\label{sec:6}Discussion and conclusions}

The gravitational lensing effect has become an important tool for studying the spacetime structure of BHs. In particular, the 2019 image of M87* revealed its high consistency with the Kerr metric, strongly supporting the theoretical expectation that black holes follow the Kerr metric. However, the existence of accretion flows, dark matter halos, and non-vacuum environments poses significant challenges to the isolated Kerr BH model. As a natural extension of the Kerr solution, the rotating short-haired BH model may provide a new theoretical framework for studying real BHs in the universe and their complex environments.

This paper investigates the strong gravitational lensing effect in the rotating short-haired BH and conducts a constraint analysis on relevant parameters. We find that the hair parameter significantly influences the black hole shadow and the gravitational lensing effect. Specifically, as the intensity of the short hair increases, the event  horizon radius of the rotating short-haired BH decreases, and the photon   orbit radius and the impact parameter decrease with the increase of the black  hole spin. Meanwhile, the photon orbit radius and the impact parameter of the Kerr BH are larger than those of the rotating short-haired BHe, and the degree of deviation increases with the increase of the  hair parameter.
For the lensing coefficients $\bar{a}$ and $\bar{b}$, they exhibit monotonically increasing and decreasing trends respectively. The hair parameter intensifies this variation, while parameter $k$ suppresses the influence effect of the hair parameter.

We further analyzed the observational manifestations of the rotating short-haired BH in M87* and Sgr A*. Whether for M87* or Sgr A*, the changing trends of these observables are consistent: the angular position of the relativistic image \(\theta_\infty\) decreases as the spin parameter increases, the angular separation \(S\) increases with the increase of the spin parameter, and the magnification of the relativistic image \(r_{mag}\) decreases as the spin parameter increases. 
The existence of the hair parameter $Q_m$ causes both $\theta_\infty$ and $S$ to be smaller than in the Kerr BH case, while $r_{mag}$ is larger than in the Kerr BH case. As the hair parameter $Q_m$ increases, this deviation gradually strengthens; however, an increase in parameter $k$ suppresses this effect, causing these observables to gradually approach the Kerr BH case.
By analyzing the data in Table \ref{table2}, it can be concluded that the hair parameter \(Q_m\) has a remarkable impact on the angular position of the relativistic image. Its variation ranges are \(22.7113\ \mu as>\theta_\infty(M87^*)>13.9162\ \mu as\) and \(29.3132\ \mu as>\theta_\infty(Sgr A^*)>17.9614\ \mu as\) respectively. These ranges are comparable to the observational results of the shadow diameters of M87* and Sgr A* obtained by the EHT \cite{EventHorizonTelescope:2019dse,EventHorizonTelescope:2022xqj} (see Figure \ref{l} and Table \ref{table2}).
Through comparing the rotating short-haired BH with the Kerr BH (\(Q_m = 0\)) and the Kerr-Newman BH (\(k = 1\)), we've found that the angular position deviation between the rotating short-haired BH and the Kerr BH can reach \(\delta\theta_\infty\approx3\ \mu as\) in the case of M87*, and \(\delta\theta_\infty\approx3.3786\ \mu as\) for Sgr A*. The deviations from the Kerr-Newman BH are \(\delta\theta_\infty\approx2.4249\ \mu as\) (for M87*) and \(\delta\theta_\infty\approx3.1299\ \mu as\) (for Sgr A*), respectively.

In addition, the time  delay effect of the rotating short-haired BH also shows significant differences from other models. When simulating M87*, the time deviation between the rotating short-haired BH and the Kerr BH can reach more than $50$ hours, while the deviation from the Kerr-Newman BH can reach more than $100$ hours. For Sgr A*, the time deviations are $2$ minutes and $4$ minutes respectively. Apparently, if the rotating short-haired BH is taken as the M87* model, the time delay effect is sufficient to distinguish it from the Kerr BH or the Kerr-Newman BH, providing theoretical support for further exploration of the properties of the hair parameter. It is worth mentioning that the maximum value of the time  delay deviation does not increase monotonically with the spin parameter, but reaches its maximum at a specific spin parameter (see Figures \ref{fig8} and \ref{fig9} and Table \ref{table4}).

Finally, by integrating the observational data from the EHT, we systematically constrained the hair parameter of the rotating short-haired BH. The results indicate that, within the first confidence interval, the parameter space \((Q_m/M, a/M)\) of the rotating short-haired BH is highly consistent with the observational data (Figure \ref{fig10}). This suggests that the rotating short-haired BH cannot be excluded as a possible model for BHs in the real universe, providing theoretical support for further differentiating between different BH models. 

In summary, the hair parameter has a significant impact on the strong gravitational lensing effect and the black hole shadow. These impacts may lead to certain observational differences for distinguishing between different BH models. Future research could further explore the influence of the hair parameter on the weak gravitational lensing effect and further verify the applicability of this model through high precision observations.

\section{acknowledgements}
We acknowledge the anonymous referee for a constructive report that has significantly improved this paper.This work was supported by Guizhou Provincial Basic Research Program(Natural Science)(Grant No.QianKeHeJiChu-[2024]Young166), the Special Natural Science Fund of Guizhou University (Grant No.X2022133), the National Natural Science Foundation of China (Grant No.12365008) and the Guizhou Provincial Basic Research Program (Natural Science) (Grant No.QianKeHeJiChu-ZK[2024]YiBan027 and QianKeHeJiChu-MS[2025]680). 

\bibliography{ref}
\bibliographystyle{apsrev4-1}

\end{document}